\documentclass[12pt,twoside]{article}
\usepackage{amssymb}
\usepackage{amsmath}
\usepackage{latexsym}
\usepackage{longtable}
\usepackage{epsfig}
\usepackage[english]{babel}
\usepackage[latin1]{inputenc}
\usepackage[T1]{fontenc}
\usepackage{ae,aecompl}
\usepackage{times}
\usepackage{amsfonts,bbm,theorem}
\usepackage{amsmath}
\newenvironment{Proof}{{\bf Proof:}}{\hspace*{\fill}$\Box$}

\newcounter{jlisti}

\newcommand{\sabs}[1]{\langle #1\rangle}

\newcommand{\Msing}{M^{\rm sing}}
\newcommand{\Fbcr}{F^{\rm cr}}

\newcommand{\R}{{\mathbb R}}

\newcommand{\T}{{\mathbb T}}
\newcommand{\rme}{{\rm e}}

 \newcommand{\ci}{{\rm i}}

\begin{document}
\vspace*{2cm}
\begin{center}
{\Large {\bf Not to normal order -- Notes on the kinetic limit for
weakly interacting quantum fluids\bigskip\\}} {\large{Jani
Lukkarinen$^\star$ and 
Herbert Spohn$^\dag$}}\bigskip\bigskip\\
 {Department of Mathematics and Statistics, University of Helsinki,\\
 P.O.\ Box 68, FI-00014 Helsingin yliopisto, Finland\\
 Zentrum Mathematik, TU M\"unchen,
 D-85747 Garching, Germany\\
 $^\star$ e-mail:~jani.lukkarinen@helsinki.fi}\\
{$^\dag$ e-mail:~spohn@ma.tum.de}
\end{center}\bigskip\bigskip
\textit{J\"{u}rg Fr\"{o}hlich and Tom
Spencer have profoundly shaped the interface between physics and mathematics.
It is therefore a particular pleasure to dedicate to them this article on a 
chapter of mathematical physics which is far from being closed.}
\bigskip\bigskip\bigskip\\
\textbf{Abstract.}
The derivation of the Nordheim-Boltzmann transport equation for
weakly interacting quantum fluids is a longstanding problem in
mathematical physics. Inspired by the method developed to handle classical
dilute gases, a conventional approach is the use of the BBGKY hierarchy
for the time-dependent reduced density matrices. In contrast, our
contribution is motivated by the kinetic theory of the weakly nonlinear
Schr\"{o}dinger equation. The main observation is that the results
obtained in the latter context carry over directly to weakly
interacting quantum fluids provided one does not insist on normal
order in the Duhamel expansion. We discuss the term by term convergence
of the expansion and the equilibrium time correlation $\langle
a(t)^\ast a(0)\rangle$.

\newpage

%%%%%%%%%%%%%%%%%%%%%%%%%%%%%%%%%%%%%%%%%%%%%%%%%%%%%%%%%%%%%%%

\section{Introduction}\label{sec.1}
\setcounter{equation}{0}

With the discovery of quantum mechanics, evidently, Boltzmann's
kinetic theory of rarified gases had to be revised. The modification
turned out to be minimal, in a certain sense. Only the classical
differential scattering cross section had to be replaced by its
quantum version, which thus depends on whether the gas particles are
fermions or bosons. Otherwise the structure of the equation remains
unaltered. In particular, the stationary solutions are still the
Maxwellians. In addition, a natural task was to investigate the kinetic 
regime for weakly interacting quantum fluids with no particular 
restriction on the density. Time-dependent perturbation theory, 
or even the more basic Fermi's golden rule, provides a convenient tool.
Since the interaction is weak, the differential cross section appears 
now only in the Born approximation. The resulting
quantum kinetic equation
has a cubic collision operator, rather than a quadratic one as in
Boltzmann's work. As a consequence, the Maxwellians
$\exp[-\beta(\omega(k)-\mu)]$ are no longer stationary. The correct
stationary solutions are of the form $(\exp[\beta(\omega(k)-\mu)]\mp
1)^{-1}$ in accordance with the statistics of the quantum fluid
under consideration.

The quantum transport equation was first written down by Nordheim
\cite{N29} in a paper submitted on May 30, 1928 and published a few
weeks later. His work is an ingenious guess, supported by an
H-theorem and by the physically expected stationary momentum
distributions. Later on more systematic derivations followed
\cite{P29}. In the kinetic literature the transport equation mostly
carries the names of Uehling and Uhlenbeck \cite{UU33}. In their
1933 paper they study properties of the transport equation, in
particular its linearization around equilibrium and the long time
hydrodynamic approximation.

There is a fundamental difference between weakly interacting 
classical and quantum fluids in the kinetic regime. For two
classical particles a small interaction results in a small change 
of the relative momentum. Therefore the collision operator is not 
an integral operator, as in Boltzmann's classic work, but it is
a nonlinear differential operator, apparently 
first realized in 1936 by Landau \cite{La36} in the context of
plasmas and fluids with long range interactions.
On the other hand, two weakly interacting quantum particles will 
pass through each other with probability $1- \mathcal{O}(\lambda^2)$,
$\lambda$ the coupling constant, and will s-wave scatter with probability
$\mathcal{O}(\lambda^2)$. Therefore the collision operator is still an
integral operator. 

In our notes we will consider only quantum fluids interacting
through a weak pair potential.

The conventional formal derivation of the transport equation
proceeds via Fermi's golden rule. Over the years there have been
many attempts to improve on the argument. On the theoretical side we
mention in particular van Hove \cite{vH}, Prigogine \cite{Pr}, and
Hugenholtz \cite{Hu}. With the work of Lanford \cite{La75} on the
microscopic justification of the classical Boltzmann equation and
the work of Davies \cite{Da76} on the weak coupling limit for a
small quantum system coupled to a large heat bath, the derivation of
kinetic equations was recognized as a problem of interest to
mathematical physics. In fact, it is already stated verbally in
Hilbert's famous collection of problems as problem No.\ 6 \cite{Hil}.
The derivation of quantum kinetic equations still stands as a
challenge. We will explain its current status in due course.

Regarding a quantum fluid as starting point, there are two in
essence orthogonal ways to proceed with a semiclassical
approximation. Within the particle picture, it is natural to take the
limit of a large mass which leads to classical point particles
interacting through a pair potential. On the other side, from the
point of view of operator-valued fields, the natural limit is a
weak, long range (on the scale of a typical inter-particle distance)
potential which leads to the nonlinear Schr\"{o}dinger equation,
also known as Hartree equation. Both limits can be formulated as an
Egorov theorem for operators \cite{Rob,Fr,FrKn,ErSch}. In the former
case one is back to the model studied by Boltzmann while in the
latter case the kinetic issue concerns a weakly nonlinear wave
equation as already studied by Peierls \cite{P29}.

In \cite{LuSp} we investigate the weakly nonlinear Schr\"{o}dinger
equation and develop a machinery for dealing with the
high-dimensional oscillatory integrals as they arise in the Duhamel
expansion of the solution of that equation. The goal of this paper
is to explain how these novel techniques can be used for the
derivation of quantum kinetic equations. In fact, we will provide
mostly sketches of the argument and prove only a few crucial points.
Otherwise we would be overwhelmed by technical issues. But we do
clearly indicate where, to our understanding, there are gaps and new
ideas will be required.

To give a brief outline: Weakly interacting lattice bosons and
lattice fermions are introduced in Section \ref{sec.2} together with
the Duhamel expansion of the time evolution operator. In Section
\ref{sec.3} we explain how this expansion differs from the expansion
resulting from the quantum BBGKY hierarchy and in Section 
\ref{sec.4} we make a comparison with wave turbulence for the weakly nonlinear
Schr\"{o}dinger equation.  The main body of novel
results are in Sections \ref{sec.6} to \ref{sec.7} where we discuss
equilibrium time correlations and spatially homogeneous
nonequilibrium states in the kinetic limit. In an appendix we
summarize a few properties of spatially homogeneous quantum kinetic
equations.\medskip\\
{\bf Acknowledgments.} We would like to thank L\'{a}szl\'{o} Erd\H{o}s for
many illuminating discussions on the subject.  
The research was supported
by a DFG grant and by the Academy of Finland.
The work was partially done at the 
the Erwin
Schr\"{o}din\-ger Institute for Mathematical Physics (ESI), 
Vienna, Austria, whom we thank for their hospitality.

%%%%%%%%%%%%%%%%%%%%%%%%%%%%%%%%%%%%%%%%%%%%%%%%%%%%%%%%%%%%%%%%%%%

\section{Weakly interacting quantum fluids, Duhamel expansion}\label{sec.2}
\setcounter{equation}{0}
\setcounter{figure}{0}

We consider quantum particles on the $d$-dimensional lattice
$\mathbb{Z}^d$ with lattice points $x\in\mathbb{Z}^d$. Later on
there will be conditions which require $d\geq 3$. The particles ``hop''
independently to nearby sites. The hopping amplitude for the
relative displacement is given by $\alpha:\mathbb{Z}^d\to
\mathbb{R}$ with $\alpha(x)=\alpha(-x)$. $\alpha$ is of compact
support, i.e., $\alpha(x)=0$ for $|x|>R$ with suitable $R$. The
Fourier transform of $\alpha$ is the dispersion relation
$\omega(k)=\widehat{\alpha}(k)$. Clearly
$\omega(k)=\omega(k)^\ast=\omega(-k)$ and $\omega$ is real analytic.
This by itself will not be enough and further conditions,
originating from the analysis in \cite{LuSp}, will be imposed. The
particles interact through the weak pair potential $\lambda V$,
$V:\mathbb{Z}^d\to \mathbb{R}$, $V(x)=V(-x)$. $V$ is assumed to be
of compact support and $\lambda>0$, $\lambda\ll 1$. Particles are
either bosons or fermions. We introduce the corresponding
annihilation/creation operators $a(x)$, $a(x)^\ast$,
$x\in\mathbb{Z}^d$. In the case of bosons they satisfy the
commutation relations
\begin{equation}\label{2.1}
[a(x),a(y)^\ast]=\delta_{xy}\,,\quad [a(x),a(y)]=0\,,\quad
[a(x)^\ast,a(y)^\ast]=0\,,
\end{equation}
while in the case of fermions they satisfy the anticommutation relations
\begin{equation}\label{2.2}
\{a(x),a(y)^\ast\}=\delta_{xy}\,,\quad \{a(x),a(y)\}=0\,,\quad
\{a(x)^\ast,a(y)^\ast\}=0\,,
\end{equation}
$x,y\in \mathbb{Z}^d$. Here $[A,B]= AB-BA$ and $\{A,B\}= AB+BA$.
With this notation the Hamiltonian of our system of particles reads
\begin{equation}\label{2.3}
H=\sum_{x,y\in\mathbb{Z}^d}\alpha(x-y)a(x)^\ast
a(y)+\tfrac{1}{2}\lambda \sum_{x,y\in\mathbb{Z}^d} V(x-y) a(x)^\ast
a(y)^\ast a(y)a(x)\,.
\end{equation}
$H$ at $\lambda=0$ is quadratic and denoted by $H_\mathrm{har}$.
Clearly the number of particles, $N$, is conserved,
\begin{equation}\label{2.3a}
N=\sum_{x\in\mathbb{Z}^d} a(x)^\ast a(x)\quad \textrm{and}\quad
[H,N]=0\,.
\end{equation}

For $f:\mathbb{Z}^d\to \mathbb{C}$ we introduce its Fourier
transform as
\begin{equation}\label{2.4}
\hat{f}(k)=\sum_{x\in\mathbb{Z}^d}f(x)\mathrm{e}^{-\mathrm{i}
2\pi k\cdot x}\,.
\end{equation}
with $k \in \mathbb{T}^d = [- \tfrac{1}{2}, \tfrac{1}{2}]^d$, the
$d$-dimensional unit torus. The physical momentum is $2\pi k$, but
our convention has the advantage of minimizing the number of $2\pi$ prefactors. The
inverse transform to (\ref{2.4}) reads
\begin{equation}\label{2.5}
f(x)=\int_{\mathbb{T}^d}dk \hat{f}(k)\mathrm{e}^{\mathrm{i} 2\pi
k\cdot x}\,.
\end{equation}
Sometimes it is more convenient to have a function, say $W$, defined
on $\mathbb{T}^d$. Then its inverse Fourier transform will be
denoted by $\check{W}$. With this notation we have in momentum space
\begin{eqnarray}\label{2.6}
&&\hspace{-20pt} H=\int_{\mathbb{T}^d}dk\,
\omega(k)\hat{a}(k)^\ast\hat{a}(k)
+\frac{1}{2}\lambda\int_{(\mathbb{T}^d)^4}dk_1 dk_2 dk_3 dk_4
\delta(k_1+k_2-k_3-k_4)\nonumber\\
&&\hspace{70pt}\times\widehat{V}(k_2-k_3)
\hat{a}(k_1)^\ast\hat{a}(k_2)^\ast
\hat{a}(k_3)\hat{a}(k_4)\,.\medskip
\end{eqnarray}
\textit{Remark 2.1 (Infinite volume limit).} The Hamiltonians
(\ref{2.3}), resp.\ (\ref{2.6}), define a unitary dynamics for
initial states which have a bounded number of particles. To be more precise,
we introduce the Fock space
\begin{equation}\label{2.7b}
\mathfrak{F}=\bigoplus^\infty_{n=0} \mathfrak{F}_n\,,\quad
\mathfrak{F}_n=\ell_2(\mathbb{Z}^d)^{\otimes n}\,.
\end{equation}
and the projections $P_\theta$, $\theta = \pm$. $P_+\mathfrak{F}$ is
the subspace of symmetric and $P_-\mathfrak{F}$ is the subspace of
antisymmetric wave functions. Since $[H,N]=0$,
$P_\theta\mathfrak{F}_n$ is invariant and $H$ restricted to
$P_\theta\mathfrak{F}_n$ is a self-adjoint and bounded operator for
each $n$. Of particular interest will be translation invariant
initial states. Then the number of particles is infinite and the
expression (\ref{2.3}) is only formal. Thus one first has to
restrict the fluid to a box $\Lambda$ with periodic boundary
conditions. In the kinetic limit its side-length must be of order
$\lambda^{-2}$. Thus, for constant density, the average number of
particles is finite and of order $\lambda^{-2d}$. Hence, for fixed
$\lambda$ the dynamics is well defined. Conceptually one would
prefer to first let $\Lambda\uparrow\mathbb{Z}^d$ and then
$\lambda\to 0$ together with the appropriate rescaling of
space-time. This program can be carried through for fermions, in
which case the operators $a(x)$, $x\in\mathbb{Z}^d$, are bounded and
the dynamics is defined as a group of automorphisms on the
$C^\ast$-algebra of quasi-local observables \cite{Ro}. Thus one can
work directly at infinite volume. For lattice bosons the dynamics at
infinite volume is not so well understood. For our purposes the
infinite volume limit is not a central issue. One can first write
down the Duhamel expansion in finite volume $\Lambda$ and then
establish that, for the particular initial state and the observables
of interest, the estimates are uniform in $\Lambda$, hence hold when
$\Lambda\uparrow\mathbb{Z}^d$.
\hspace*{0cm}\hfill{$\diamondsuit$}\medskip\\
\textit{Remark 2.2 (Stability)}. We will study equilibrium time
correlations in the kinetic limit. For this the $\beta$-KMS state
$\langle\cdot\rangle_{\beta,\lambda}$ for $H-\mu N$, $\mu$ the
chemical potential, has to be well-defined and, hence, $V$ has to be
a thermodynamically stable potential. For lattice fermions no
further conditions are needed. For lattice bosons a simple
sufficient condition would be
\begin{equation}\label{2.7aa}
V\geq 0\,.
\end{equation}
For a discussion of equilibrium states at small $\lambda$ we refer
to \cite{Gi,BrRo}. In the kinetic limit thermodynamic stability does
seem to play a role.
 \hspace*{0cm}\hfill{$\diamondsuit$}\medskip\\
\textit{Remark 2.3 (Continuum limit).} In \cite{LuSp} we study a
classical field theory by regarding $a(x),a(x)^\ast$ as commutative
complex-valued field. To avoid ultraviolet divergence, it was
necessary to introduce a spatial discretization, as will be
explained in more detail in Section \ref{sec.4}. Since we plan to
transcribe the results from \cite{LuSp} to quantum fluids, we stick
to the lattice theory. Thus our model is appropriate, for example,
for a fluid of electrons in a crystal background potential and for
bosons in an optical lattice. To understand whether a continuum
limit is feasible, we recall that
\begin{equation}\label{2.7}
\langle\hat{a}(k)^\ast \hat{a}(k')\rangle_{\beta,0}
=\delta(k - k') W^{\pm}_{\beta}(k)\,,\quad
W^{\pm}_{\beta}(k)=\frac{1}{\mathrm{e}^{\beta(\omega(k)-\mu)}\mp
1}\,.
\end{equation}
Here $\beta$ is the inverse temperature, $\beta>0$, and $\mu$ is the
chemical potential chosen such that $\omega(k)-\mu>0$ (and
suppressed in our notation). We adopt the standard convention
$W^{\theta}_{\beta}$, where $\theta=1$ stands for bosons and
$\theta=-1$ for fermions.

In the continuum limit, i.e., for a fluid in empty space, one would
replace the position space $\mathbb{Z}^d$ by $\mathbb{R}^d$ and
$\alpha\ast$ by $-\Delta$, implying the dispersion relation
$\omega(k)=k^2$. Clearly in this case
\begin{equation}\label{2.8}
\int_{\mathbb{R}^d}W^{\pm}_{\beta}(k) dk<\infty\,,
\end{equation}
indicating that there is no ultraviolet divergence. Indeed, as
discussed in \cite{Gi,BrRo}, the equilibrium state for small
$\lambda$, and with restrictions on the density, is well-defined. It
would be of considerable interest to find out whether the analysis
in \cite{LuSp} could be extended to weakly interacting quantum
fluids in $\mathbb{R}^d$.
\medskip\hspace*{0cm}\hfill{$\diamondsuit$}

For an arbitrary operator $A$ we define its Heisenberg evolution by
$A(t)=\mathrm{e}^{\mathrm{i}Ht}A\mathrm{e}^{-\mathrm{i}Ht}$,
$A(0)=A$. Using $\frac{d}{dt}A(t)=\mathrm{i}[H,A(t)]$, one obtains
the evolution equation
\begin{equation}\label{2.7a}
\frac{d}{dt}\; a(x,t)=-\mathrm{i}\sum_{y\in\mathbb{Z}^d} \alpha(x-y)
a(y,t)-\mathrm{i}\lambda\sum_{y\in\mathbb{Z}^d} V(x-y)a(y,t)^\ast
a(y,t)a(x,t)\,,
\end{equation}
which in momentum space becomes
\begin{eqnarray}\label{2.10}
&&\hspace{-52pt}\frac{d}{dt}\;\hat{a}(k_1,t)=-\mathrm{i}\omega(k_1)
\hat{a}(k_1,t)-\mathrm{i}\lambda\int_{(\mathbb{T}^d)^3} dk_2
dk_3 dk_4 \delta(k_1+k_2-k_3-k_4)\nonumber\\
&&\hspace{52pt}\times\widehat{V}(k_2-k_3)
\hat{a}(k_2,t)^\ast\hat{a}(k_3,t) \hat{a}(k_4,t)\,.
\end{eqnarray}

Equations (\ref{2.7a}) and (\ref{2.10}) will be solved as a Cauchy problem
given the initial, time $t=0$, state. In general one should allow
for spatial variation on the scale $\lambda^{-2}$. To keep matters
simple we will restrict ourselves to a spatially homogeneous
situation. Thus the initial state $\langle\cdot\rangle$ is assumed
to be translation invariant with good clustering properties. In
particular
\begin{equation}\label{2.9}
\langle \hat{a}(k)^\ast \hat{a}(k')\rangle=\delta(k-k')W(k)
\end{equation}
for some smooth $W$. In addition, $\langle\cdot\rangle$ has to be
gauge invariant, i.e.,
\begin{equation}\label{2.9a}
\langle \mathrm{e}^{\mathrm{i}\vartheta N} A
\mathrm{e}^{-\mathrm{i}\vartheta N}\rangle=\langle A \rangle
\end{equation}
for $\vartheta\in[0,2\pi]$.  One quantity of interest will be the
two-point function at time $t$. Since the dynamics preserves
translation and gauge invariance, it must necessarily be of the form
\begin{equation}\label{2.9b}
\langle \hat{a}(k,t)^\ast
\hat{a}(k',t)\rangle=\delta(k-k')W_\lambda (k,t)\,,
\end{equation}
$W_\lambda\geq 0$, $W_\lambda(k,0)=W(k)$, together with $\langle \hat{a}(k,t)\rangle=0$,
$\langle \hat{a}(k,t)\hat{a}(k',t)\rangle=0$.

Kinetic theory studies $W_\lambda(k,t/ \lambda^2)$ for small $\lambda$. A
special role will be played by quasifree states, which we define
first.\medskip\\
\textbf{Definition 2.4} (quasifree state). \textit{Let $\rho_1$ be a
self-adjoint operator on $\ell_2(\mathbb{Z}^d)$ with integral kernel
$\rho_1(x,y)$. For bosons we impose $\rho_1\geq 0$ while for fermions
$0\leq\rho_1\leq 1$. A gauge invariant state $\langle\cdot\rangle$
is called quasifree with correlator $\rho_1$, if for all
$x_i,y_j\in\mathbb{Z}^d,$ $i=1,...,m,j=1,...,n$, it holds}
\begin{equation}\label{2.9b2}
   \big\langle\big(\prod^m_{i=1}a(x_i)\big)^\ast\big(\prod^n_{j=1}
   a(y_j)\big)\big\rangle
    =\delta_{mn}\det{_\theta}\big(\rho_1(x_i,y_j)\big)_{i,j=1,\ldots,n}\,.\medskip
\end{equation}
Here the operators are ordered from left to right as they appear in
the $\prod$ symbol and the empty product is interpreted as 1. $\det_-$ is the usual determinant, while
$\det_+$ is the permanent. Translation invariance is reflected by
$\rho_1(x,y)$ depending only on $x-y$, resp.\ in momentum space by
(\ref{2.9}).

Kinetic theory puts forward a rather simple picture of the dynamics
for small $\lambda$: On the microscopic time scale the initial
$W(k)$ does not change while the unperturbed dynamics forces the
state to become quasifree. To say, after a time $t$ which is short
on the kinetic scale and long on the microscopic scale, for
arbitrary $n \in \mathbb{N}$, the higher moments are approximately
of the form, $\check{W}$ denoting the inverse Fourier transform,
\begin{equation}\label{2.9c}
\big\langle\big(\prod^n_{i=1}a(x_i,t)\big)^\ast
\big(\prod^n_{j=1}a(y_j,t)\big)\big\rangle\simeq
{\det}_{\theta}\big(\check{W}(x_i-y_j)\big)_{i,j=1,\ldots,n}
\end{equation}
with all other moments vanishing. On the kinetic time scale,
$t=\mathcal{O}(\lambda^{-2})$, $W_\lambda(k,t)$ changes while
preserving the quasifree property. Of course, only in the limit
$\lambda\to 0$ we obtain such a strict separation of the two
space-time scales. The initial time slip with the dynamics generated
by $H_\mathrm{har}$ was studied by Ho and Landau \cite{HoLa}. Under
suitable assumptions on $\omega$, they prove quasifreeness in the
limit $t\to\infty$ provided the initial state is
$\ell_1$-clustering.

We are mostly interested in the kinetic time scale and thus impose
the state $\langle\cdot\rangle$ to be quasifree to begin with. As
argued in \cite{ESY04}, see also \cite{Sp06}, if one assumes, up to
small errors, the state still to be quasifree at the long time
$t=\lambda^{-2}\tau$, $\tau=\mathcal{O}(1)$, then it is not too
difficult to determine the (approximate) evolution equation for
$W_\lambda(k,t)$.

To study $\langle \hat{a}(k,t)^\ast \hat{a}(k',t)\rangle$
the only method currently available is to expand the expectation
value of interest with respect to $\lambda$, which is achieved
through the Duhamel expansion. It will be convenient to work in the
interaction representation and we introduce
\begin{equation}\label{2.9d}
\hat{a}(k,1,t)=\mathrm{e}^{\mathrm{i}\omega(k)t}\hat{a}(k,t)\,,\quad
\hat{a}(k,-1,t)=\mathrm{e}^{-\mathrm{i}\omega(k)t}\hat{a}(-k,t)^\ast\,.
\end{equation}
Then (\ref{2.10}) becomes
\begin{eqnarray}\label{2.11}
&&\hspace{-23pt}\frac{d}{dt}\hat{a}(k_1,\sigma,t)=
-\mathrm{i}\lambda \sigma
 \int_{(\mathbb{T}^d)^3} dk_2 dk_3
dk_4 \delta(k_1-k_2-k_3-k_4)\nonumber\\
&&\hspace{40pt}\times
\tfrac{1}{2}\big((1+\sigma)\widehat{V}(k_2+k_3)+(1-\sigma)\widehat{V}(k_3+k_4)\big)
\nonumber\\
&&\hspace{40pt} \times\exp\big[-\mathrm{i}t\big(-\sigma \omega(k_1)-
\omega(k_2)+\sigma\omega(k_3)+\omega(k_4)\big)\big]\nonumber\\
&&\hspace{40pt}
\times \hat{a}(k_2,-1,t)
\hat{a}(k_3,\sigma,t) \hat{a}(k_4,1,t)\,.
\end{eqnarray}
For a product it holds that
\begin{equation}\label{2.12}
\frac{d}{dt}\prod^n_{j=1}\hat{a}(k_j,\sigma_j,t)=
\sum^n_{m=1}\big(\prod^{m-1}_{j=1} \hat{a}(k_j,\sigma_j,t)\big)
\frac{d}{dt}\;\hat{a}(k_m,\sigma_m,t) \big(\prod^n_{j=m+1}
\hat{a}(k_j,\sigma_j,t)\big)\,,
\end{equation}
which, using (\ref{2.11}), integrates in time to
\begin{eqnarray}\label{2.13}
&&\hspace{-30pt} \prod^n_{j=1}\hat{a}(k_j,\sigma_j,t)=
\prod^n_{j=1}\hat{a}(k_j,\sigma_j)-\mathrm{i}\lambda\sum^n_{m=1}\sigma_m\int^t_0
ds
\int_{(\mathbb{T}^d)^3}dk'_2 dk'_3 dk'_4\nonumber\\
&&\hspace{-10pt} \times\delta(k_m-k'_2-k'_3-k'_4)\tfrac{1}{2}\big((1+\sigma_m)
\widehat{V}(k'_2+k'_3)+(1-\sigma_m)\widehat{V}(k'_3+k'_4)\big)\nonumber\\
&&\hspace{-10pt} \times\exp\big[-\mathrm{i}s(-\sigma_m\omega(k_m)
-\omega(k'_2)+\sigma_m\omega(k'_3)+\omega(k'_4))\big]\\
&&\hspace{-10pt}\times \big(\prod^{m-1}_{j=
1}\hat{a}(k_j,\sigma_j,s)\big)\hat{a}(k'_2,-1,s)\hat{a}(k'_3,\sigma_m,s)
\hat{a}(k'_4,1,s)\big(\prod^n_{j=m+1}\hat{a}(k_j,\sigma_j,s)\big)\,.\nonumber
\end{eqnarray}
Here the products are ordered from left to right with increasing
label $j$.

\begin{figure}[t]
\setlength{\unitlength}{1cm}
\begin{picture}
(10,7.2)(-2.5,-0.5)
\put(-0.5,-0.1){$0$} \put(-0.5,6){$t$} \put(0.1,0.7){$s_0$}
\put(0.1,2.2){$s_1$} \put(0.1,3.7){$s_2$} \put(0.1,5.2){$s_3$}
 \put(8.7,0.7){$k_{0,1},...,k_{0,8}$}
 \put(8.7,2.2){$k_{1,1},...,k_{1,6}$}
\put(8.7,3.7){$k_{2,1},...,k_{2,4}$}
\put(8.7,5.2){$k_{3,1},k_{3,2}$}

\linethickness{0.1pt}
\put(0,0){\line(1,0){9}}
\put(0,1.5){\line(1,0){9}}
\put(0,3){\line(1,0){9}}
\put(0,4.5){\line(1,0){9}}
\put(0,6){\line(1,0){9}}

\thicklines
\put(4,4.5){\line(0,1){1.5}}
\put(7,4.5){\line(0,1){1.5}}
\put(2,3){\line(0,1){0.5}}
\put(4,3){\line(0,1){1.5}}
\put(5,3){\line(0,1){0.5}}
\put(7,3){\line(0,1){1.5}}
\put(2,1.5){\line(0,1){1.5}}
\put(4,1.5){\line(0,1){1.5}}
\put(5,1.5){\line(0,1){1.5}}
\put(6,1.5){\line(0,1){0.5}}
\put(7,1.5){\line(0,1){1.5}}
\put(8,1.5){\line(0,1){0.5}}
\put(1,0){\line(0,1){0.5}}
\put(2,0){\line(0,1){1.5}}
\put(3,0){\line(0,1){0.5}}
\put(4,0){\line(0,1){1.5}}
\put(5,0){\line(0,1){1.5}}
\put(6,0){\line(0,1){1.5}}
\put(7,0){\line(0,1){1.5}}
\put(8,0){\line(0,1){1.5}}
\put(8,0){\line(0,1){1.5}}
\put(3,4.5){\line(1,0){1}}

\put(2.0,0.5){\oval(2,2)[tl]}
\put(2.0,0.5){\oval(2,2)[tr]}
\put(3.0,3.5){\oval(2,2)[tl]}
\put(4.0,3.5){\oval(2,2)[tr]}
\put(7.0,2.0){\oval(2,2)[tl]}
\put(7.0,2.0){\oval(2,2)[tr]}

\put(1,0.6){\vector(0,-1){0.2}}
\put(2,0.6){\vector(0,-1){0.2}}
\put(3,0.5){\vector(0,1){0.2}}
\put(4,0.6){\vector(0,-1){0.2}}
\put(5,0.5){\vector(0,1){0.2}}
\put(6,0.6){\vector(0,-1){0.2}}
\put(7,0.5){\vector(0,1){0.2}}
\put(8,0.5){\vector(0,1){0.2}}
\put(2,2.1){\vector(0,-1){0.2}}
\put(4,5.1){\vector(0,-1){0.2}}
\put(7,5.0){\vector(0,1){0.2}}

\put(2,1.5){\circle*{0.2}}
\put(4,4.5){\circle*{0.2}}
\put(7,3){\circle*{0.2}}

\end{picture}
\caption{A Feynman diagram with $n_0 = 2$, $n=3$. Up (down) arrows
correspond to parity $+1$ ($-1$). The operator ordering at $t=0$ is
$a^\ast a^\ast a a^\ast a a^\ast aa$. The interaction history is
(1,4,1).} \label{fig1}
\end{figure}
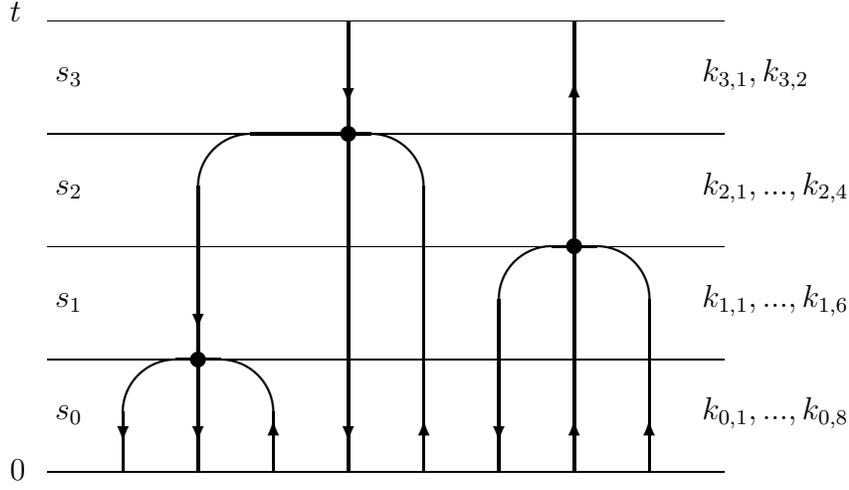

Iteration of (\ref{2.13}) yields the Duhamel expansion for products
of the form
\begin{equation}\label{2.14}
\prod^{n_0}_{j=1} \hat{a}(k_j,\sigma_j,t)\,,\quad t>0\,.
\end{equation}
The expansion is most concisely organized through Feynman diagrams,
compare with Fig.\ \ref{fig1}. We explain a generic term.

We denote the final time by $t$, $t>0$. The initial time is 0. There
will be $n$ ``collisions'' or interactions, $n\geq 0$, which in the
Feynman diagram are represented as fusions. We subdivide $[0,t]$
into $n+1$ time slices with index $i\in I_{0,n}=\{0,\ldots,n\}$.
Omitting the zero we set $I_n=\{1,\ldots,n\}$. The $i$-th slice has
length $s_i$. Thus $\sum^n_{i=0}s_i=t$ and the slices are
$[0,s_0],\ldots,[s_0+\ldots+s_{n-1},t]$. To describe the fusions we
start in slice 0 with $n_0+2n$ line segments, see Fig.\ \ref{fig1}
for the orientation of the time axis. At time $t=s_0$ exactly one
triplet of neighboring line segments fuses into a single line
segment, while all remaining line segments are continued vertically.
Thus in time slice 1 one has exactly $n_0+2n-2$ line segments. At
time $t=s_0+s_1$ exactly one triplet of neighboring line segments
fuses into a single line segment, etc.. The last fusion is at
$t=s_0+\ldots+s_{n-1}$ and in slice $n$ one has $n_0$ line segments.
Sometimes it is convenient to read the Feynman diagram backwards in
time. The only change is that ``fusion'' is turned into
``branching''. We now label the line segments by momenta $k_{i,j}\in
\mathbb{T}^d$ and by parities $\sigma_{i,j}\in \{-1,1\}$. The first
index is the label of the slice. In the $i$-th slice we label the
momenta and parities from left to right by $j=1,\ldots,m_i$ with
$m_i=n_0+2n-2i$. The corresponding index set is denoted by
$\mathcal{I}_{n;n_0}=\{(i,j)|0\leq i\leq n$, $1\leq j\leq m_i\}$. An
interaction history is denoted by $\ell=(\ell_1,\ldots,\ell_n)$,
where $\ell_i$ refers to the index of the fused line segment in slice
$i$, which is also the index of the leftmost line segment in the
fusing triplet in slice $i-1$. The set of all interaction histories
is denoted by $G_n$. Clearly $G_0=\emptyset$ and
$G_n=I_{m_1}\times\ldots\times I_{m_n}$.

With these preparations the $n$-th term of the expansion reads
\begin{eqnarray}\label{2.15}
&&\hspace{-15pt}\mathcal{F}_n
(t,k_{n,1},\ldots,k_{n,n_0},\sigma_{n,1},\ldots,\sigma_{n,n_0})
[\hat{a}]\\
&&\hspace{-0pt}=(-\mathrm{i}\lambda)^n \sum_{\ell\in G_n}
\sum_{\sigma\in\{-1,1\}^{\mathcal{I}_{n;n_0}}}
\int_{(\mathbb{T}^d)^{\mathcal{I}_{n;n_0}}} dk
\Delta_{n,\ell}(k,\sigma)\nonumber\\
&&\hspace{10pt}\times
\prod^n_{i=1}\big\{\tfrac{1}{2}(1+\sigma_{i,\ell_i})\hat{V}(k_{i-1,\ell_i}
+k_{i-1,\ell_i+1})+\tfrac{1}{2}(1-\sigma_{i,\ell_i})\hat{V}(k_{i-1,\ell_i+1}
+k_{i-1,\ell_i+2})\big\}\nonumber\\
&&\hspace{10pt}\times\prod^{n_0+2n}_{j=1}
\hat{a}(k_{0,j},\sigma_{0,j})\int_{(\mathbb{R}_+)^{I_{0,n}}}d\underline{s}
\,\delta (t-\sum^n_{i=0} s_i) \prod^n_{i=1} \exp [-\mathrm{i}
t_i(\underline{s}) \Omega_{i-1;\ell_i}(k,\sigma_{i,\ell_i})]\,\nonumber
\end{eqnarray}
as a polynomial of order $n_0+2n$ in the initial fields. Here we
have introduced the following shorthands,
\begin{equation}\label{2.16}
t_i(\underline{s})=\sum^{i-1}_{j=0}s_j\,,\quad i=1,\ldots,n\,,
\end{equation}
and
\begin{equation}\label{2.17}
\Omega_{i;j}(k,\sigma)=-\sigma\omega(k_{i,j}+k_{i,j+1}+k_{i,j+2})-
\omega(k_{i,j})+\sigma\omega(k_{i,j+1})+\omega(k_{i,j+2})\,.
\end{equation}
$\Delta_{n,\ell}$ contains the $\delta$-functions restricting the
integral over $k$ and the sum over $\sigma$ to the graph defined by
the interaction history $\ell$. As can be seen from (\ref{2.13}),
for non-fusing line segments the momentum and the parity are
transported unchanged, while the triplet of neighboring line
segments with parities $-,+,+$ fuses into a line segment with parity
$+$ and the triplet of neighboring line segments with parities $-,-,+$
fuses into a line segment with parity $-$. In other words, the middle line conserves parity, while the left neighbor carries parity $-$ and the right neighbor parity $+$. Thus, if the parities in slice $n$ are given, then they are already determined for the remaining diagram. At a fusion
the total momentum is conserved (Kirchhoff's rule). Explicitly,
\begin{eqnarray}\label{2.18}
&&\hspace{-24pt}\Delta_{n,\ell}(k,\sigma)=
\prod^n_{i=1}\Big\{\prod^{\ell_i-1}_{j=1}\big[\delta(k_{i,j}-k_{i-1,j})
\mathbbm{1}(\sigma_{i,j}=\sigma_{i-1,j})\big]
\nonumber\\
&&\hspace{30pt}\times\delta(k_{i,\ell_i}-\sum^2_{j'=0}
k_{i-1,\ell_i+j'})\mathbbm{1}(\sigma_{i-1,\ell_i}=-1)
\mathbbm{1}(\sigma_{i-1,\ell_{i+1}}=\sigma_{i,\ell_i})
\mathbbm{1}(\sigma_{i-1,\ell_{i+2}}=1)\nonumber\\
&&\hspace{30pt} \times
\prod^{m_i}_{j=\ell_i+1}\big[\delta(k_{i,j}-k_{i-1,j+2})
\mathbbm{1}(\sigma_{i,j}=\sigma_{i-1,j+2})\big]\Big\}\,,
\end{eqnarray}
with $\mathbbm{1}$ the symbol for the indicator function.

The Duhamel expansion of the product (\ref{2.14}) cut at order $N$
is given by
\begin{eqnarray}\label{2.20}
&&\hspace{-55pt} \prod^{n_0}_{j=1}
\hat{a}(k_j,\sigma_j,t)=\sum^{N-1}_{n=0} \mathcal{F}_n
(t,k_1,\ldots,k_{n_0},\sigma_1,\ldots,\sigma_{n_0})[\hat{a}]
\nonumber\\
&&\hspace{30pt} +\int^t_0 ds \mathcal{F}_N
(t-s,k_1,\ldots,k_{n_0},\sigma_1,\ldots,\sigma_{n_0})
\big[\hat{a}(s)\big]\,.
\end{eqnarray}
As before, in (\ref{2.20}) $\hat{a}$ denotes the time 0 field, while
$\hat{a}(s)$ is the time $s$ field as it appears in (\ref{2.13}).
To complete the
Feynman diagrams the expansion in (\ref{2.20}) has still to be averaged
over some initial state.
Note (i) the error term still contains the full time evolution and
(ii) the operator order for the product appearing in $\mathcal{F}_n$
depends on the particular interaction history. 

Given an initial quasifree state of the form (\ref{2.9b2}) with
bounded $W$, one can estimate roughly the magnitude of a Feynman
diagram with respect to this initial state. As an example, let us
consider $\langle \hat{a}(k,t)^\ast \hat{a}(k',t)\rangle$ for which
$n_0=2$. At order $n$ of the expansion, from the momentum and time
integrations one obtains the bound $\lambda^n t^n c^n/n!$ with a
suitable constant $c$. The initial state has $n!$ terms of the same
size. Thus a Feynman diagram is bounded by $\lambda^n t^n c^n$.
However the number of collision histories is also $n!$, thus
yielding a \textit{zero} radius of convergence for the infinite
series.

One possibility to improve the situation would be to extract some
extra decay in time. A point in case is a Fermi fluid where
particles interact only if they are inside a prescribed bounded
region. Outside this region they do not interact which generates a
controllable decay. As proved in \cite{BoMa,Fr1} the Duhamel
expansion converges in operator norm provided the coupling strength
satisfies $|\lambda|<\lambda_0$ with small $\lambda_0$.

For the kinetic limit no such improved time decay seems to be
available and one is left with the following tentative program.\smallskip\\
1.) One studies the convergence of each Feynman diagram for rescaled
time $\lambda^{-2}t$, $t>0$, in the limit $\lambda\to
0$.\smallskip\\
2.) Following the pioneering work of Erd\"{o}s and Yau \cite{EY00} in case of
 the linear Schr\"{o}dinger equation with a weak random potential, one cuts the 
series at some $\lambda$-dependent $N$, the generic
choice being $\lambda N!=1$, or perhaps $\lambda^\kappa N!=1$ with a
suitable choice of $\kappa>0$, and tries to control the error term
by some other means. In combination with 1.) this would allow one to
determine the two-point function
\begin{equation}\label{2.21}
\langle \hat{a}(k,\lambda^{-2}t)^\ast
\hat{a}(k',\lambda^{-2}t)\rangle\quad \textrm{as }\lambda\to
0\,.
\end{equation}

We will argue that step 1 can be carried out under specified
conditions on the dispersion relation $\omega$. For step 2, at
present our only other help is stationarity. This means that we take
as initial state $\langle\cdot\rangle$ the KMS state for $H$ with a
suitable choice of the chemical potential. Of course, (\ref{2.21})
is then time-independent. Quantities of interest would be two-time
correlations in equilibrium as $\langle
\hat{a}(k,\lambda^{-2}t)^\ast \hat{a}(k')\rangle$ and, considerably
more difficult to handle, the number density time correlation
$\langle \hat{a}(k,\lambda^{-2}t)^\ast
\hat{a}(k,\lambda^{-2}t)\hat{a}(k')^\ast \hat{a}(k')\rangle$.

%%%%%%%%%%%%%%%%%%%%%%%%%%%%%%%%%%%%%%%%%%%%%%%%%%%%%%%%%%%%%%

\section{The quantum BBGKY hierarchy}\label{sec.3}
\setcounter{equation}{0}

We return to position space. The  $n$-th reduced density matrix, at time $t$
is given by the expectation
\begin{equation}\label{3.0}
\rho_n(x_1,y_1,\ldots,x_n,y_n,t)=\langle
\big(\prod^n_{i=1}a(y_i,t)^\ast\big)\big(\prod^1_{j=n}a(x_j,t)\big)\rangle\,.
\end{equation}
Here $\rho_n(t)$ should be regarded as the kernel of a positive
operator acting on $\ell_2(\mathbb{Z}^d)^{\otimes n}$ via summation over $y$.

For both fermions and bosons $\rho_n(t)$ is symmetric in the
arguments $(x_j,y_j)$, $j=1,\ldots,n$. In addition, for fermions
$\rho_n(t)$ is separately antisymmetric in the arguments
$(x_1,\ldots,x_n)$ and $(y_1,\ldots,y_n)$ , while it is symmetric
for bosons. To differentiate $\rho_n(t)$ in time we use
(\ref{2.7a}). The resulting operator ordering is no longer normal,
in general, and one has to normal order so as to get a closed evolution
equation for the $\rho_n(t)$'s. Let us define, as operators on
$\ell_2(\mathbb{Z}^d)^{\otimes n}$,
\begin{equation}\label{3.1}
H^{(n)}=H^{(n)}_0 + \lambda V^{(n)}\,,
\end{equation}
with
\begin{equation}\label{3.2}
H^{(n)}_0 \psi_n(x_1,\ldots,x_n)=\sum^n_{j=1}
\sum_{w\in\mathbb{Z}^d} \alpha(x_j-w)\psi_n(x_1,\ldots,w,\ldots,x_n)
\end{equation}
and
\begin{equation}\label{3.3}
V^{(n)}\psi_n(x_1,\ldots,x_n)=\tfrac{1}{2}\sum^n_{i\neq j=1}
V(x_i-x_j)\psi_n(x_1,\ldots,x_n)\,.
\end{equation}
We also define
\begin{equation}\label{3.4}
(C_{n,n+1}\rho_{n+1})(x_1,\ldots,y_n)=-\mathrm{i} \sum^n_{j=1}
\sum_{w\in\mathbb{Z}^d} \big(V(x_j-w)-V(y_j-w)\big)
\rho_{n+1}(x_1,\ldots,y_n,w,w)\,.
\end{equation}
Then the reduced density matrices satisfy
\begin{equation}\label{3.6}
\frac{d}{dt}\rho_n(t)=-\mathrm{i}[H^{(n)},\rho_n(t)]+\lambda
C_{n,n+1}\rho_{n+1}(t)\,,
\end{equation}
which is the quantum BBGKY hierarchy.

On purely mathematical grounds, the unitary evolution extends
naturally to the full Fock space $\mathfrak{F}$. A state on
$\mathfrak{F}$ is given through some positive density matrix $S$
with $\mathrm{tr}_{\mathfrak{F}}S=1$. We require gauge invariance as 
$[S,e^{i\vartheta N}]=0$ for all $\vartheta \in [0,2\pi]$ with $N$ the number operator on
$\mathfrak{F}$. Thus, denoting by $P_n$ the projection onto the $n$-particle subspace
$\mathfrak{F}_n$, it holds
\begin{equation}\label{3.6a}
P_m S P_n=\delta_{mn} S_n\,.
\end{equation}
$S_n$ is a positive operator on $\mathfrak{F}_n$ and
$\mathrm{tr}_{\mathfrak{F}_n} S_n\leq 1$. Since particles are taken
to be indistinguishable, we require $S_n$ to be permutation
invariant as an operator on $\mathfrak{F}_n$. Let us denote by
$\mathrm{tr}_{[m,n]}$, $m\leq n$, the partial trace on
$\mathfrak{F}_n$ over the tensor product factors with labels $m$ to
$n$, in particular
$\mathrm{tr}_{[1,n]}=\mathrm{tr}_{\mathfrak{F}_n}$. In this general
context the $n$-th reduced density matrix is defined through
\begin{equation}\label{3.6b}
\rho_n= \sum^\infty_{m=0}\frac{(n+m)!}{m!}\mathrm{tr}_{[n+1,n+m]} S_{n+m}
\end{equation}
as a positive operator on $\mathfrak{F}_n$, with the convention
$\mathrm{tr}_{[n+1,n]} S_n=S_n$. The normalization is such that
\begin{equation}\label{3.6c}
\mathrm{tr}_{[1,n]}\rho_n=\mathrm{tr}_\mathfrak{F}[S N(N-1)\ldots
(N-n+1)]\,,
\end{equation}
which is assumed to be bounded by $c^n$. Then (\ref{3.6b}) can be
inverted as
\begin{equation}\label{3.6c1}
S_n= \sum^\infty_{m=0}\frac{(-1)^m}{n!m!}\mathrm{tr}_{[n+1,n+m]}
\rho_{n+m}\,.
\end{equation}
If $\mathrm{tr}_{\mathfrak{F}}[P_\theta S]=1$, $\theta=\pm 1$, then, using the
realization of $a(x)$, $a(x)^\ast$ on Fock space, it is easy to
check that (\ref{3.6b}) agrees with the definition (\ref{3.0}).

The density matrix $S$ evolves in time through
\begin{equation}\label{3.6d}
P_m S(t) P_n=\delta_{mn} \mathrm{e}^{-\mathrm{i}
H^{(n)}t}S_n \mathrm{e}^{\mathrm{i} H^{(n)}t}\,.
\end{equation}
We insert in the definition (\ref{3.6b}) and differentiate with
respect to $t$. The permutation symmetry of the state is preserved
in time. This leads to
\begin{equation}\label{3.6e}
\frac{d}{dt}\rho_n(t)=-\mathrm{i}[H^{(n)},\rho_n(t)]-\mathrm{i}\lambda\sum^n_{j=1}
\mathrm{tr}_{[n+1,n+1]}\big([V_{j,n+1},\rho_{n+1}(t)]\big)\,,
\end{equation}
where $V_{j,n+1}$ is multiplication by $V(x_j-x_{n+1})$ as an
operator on $\mathfrak{F}_{n+1}$. Clearly, (\ref{3.6e}) is identical
to (\ref{3.6}).

The time-integrated version of (\ref{3.6}) reads
\begin{equation}\label{3.7}
\rho_n(t)=\mathrm{e}^{-\mathrm{i}H^{(n)}t}\rho_n\mathrm{
e}^{\mathrm{i}H^{(n)}t}+ \lambda\int^t_0 ds
\mathrm{e}^{-\mathrm{i}H^{(n)}(t-s)}C_{n,n+1}
\rho_{n+1}(s)\mathrm{e}^{\mathrm{i}H^{(n)}(t-s)}\,.
\end{equation}
Of interest is $\rho_1(t)$. Its perturbation series is generated by
iterating (\ref{3.7}). This will not be an expansion in $\lambda$,
since $H^{(n)}$ depends itself on $\lambda$. Expanding
$\exp[-\mathrm{i}t(H^{(n)}_0+\lambda V^{(n)})]$ with respect to
$\lambda$ and inserting in (\ref{3.7}) yields a perturbation
expansion in $\lambda$. For bosons and fermions it has to be in
one-to-one correspondence with the Feynman diagrams of Section 2. By
construction, for the BBGKY hierarchy the expectation with respect
to the initial state is over a normal ordered product of operators,
while for Feynman diagrams normal order does not hold, in general,
compare with (\ref{2.15}). On the other hand, reading Feynman
diagrams backwards in time, single line segments can branch only
into three line segments, while the perturbation expansion of the
BBGKY hierarchy in the form (\ref{3.7}) consists of $n$ particles
interacting amongst themselves and one extra particle added through
the ``collision'' $C_{n,n+1}$.

The BBGKY hierarchy is used by Benedetto \textit{et al.}
\cite{BCEP04,BCEP05,BCEP07} in their study of the kinetic limit. One
observes that for the free evolution generated by $H_0^{(n)}$ the
kinetic limit, space $ \sim \lambda ^{-2}$, time $ \sim \lambda
^{-2}$, is equivalent to the semiclassical limit. This can be
exploited by transforming $\rho_n$ in each of its variables
$(x_j,y_j)$ to a Wigner function. Thereby the BBGKY hierarchy turns
into a hierarchy of multi-point Wigner functions. The free part
corresponds to classical particles with kinetic energy $\omega(k)$
and the difficulty resides in handling the nonlocal ``collisions''.
Benedetto \textit{et al.} work in the continuum, $\mathbb{R}^3$
instead of $\mathbb{Z}^3$, and use the quadratic dispersion law
$\omega(k)=k^2$. The initial reduced density matrices are assumed to be
of the factorized form 
\begin{equation}\label{3.8}
\rho_n(x_1,y_1,\ldots,x_n,y_n)=\prod^n_{j=1}\rho_1(x_j,y_j)\,,
\end{equation}
at least asymptotically for small $\lambda$. (\ref{3.8}) does not
have the antisymmetry required for fermions. For bosons the equality
$\rho_2=\rho_1\otimes\rho_1$ forces $\rho_1$ to be a pure state as can be
seen from\\
\textit{Remark 3.1 (Factorization).} Using the spectral
representation of $\rho_1$ with eigenfunctions $\phi_j$ and
eigenvalues $\lambda_j\geq 0$, symmetry in $x_1,x_2$ implies
\begin{equation}\label{3.8a}
\sum^\infty_{i,j=1}\lambda_i\lambda_j
\phi_i(x_1)\phi_i(y_1)^\ast\phi_j(x_2)\phi_j(y_2)^\ast
=\sum^\infty_{i,j=1}\lambda_i\lambda_j
\phi_j(x_1)\phi_i(y_1)^\ast\phi_i(x_2)\phi_j(y_2)^\ast\,.
\end{equation}
Taking the inner product with $\phi_\alpha(x_2)$ from left and right
yields $\lambda_\alpha(1-\lambda_\alpha)=0$.
\hspace*{0cm}\hfill{$\diamondsuit$}\medskip\\

$\rho_1$ defines the scaled Wigner function through
\begin{equation}\label{3.9}
W_\lambda(r,v,t)=\int d\eta\, \mathrm{e}^{\mathrm{i}\eta\cdot
v}\rho_1(\lambda^{-2} r +\tfrac{1}{2}\eta,\lambda^{-2}
r-\tfrac{1}{2}\eta,t)\,.
\end{equation}
It is assumed that, for $t=0$,
\begin{equation}\label{3.10}
\lim_{\lambda\to 0} W_\lambda(r,v)=W_0(r,v)
\end{equation}
with $W_0(r,v)$ sufficiently smooth and of rapid decay in both
arguments. Benedetto \textit{et al.} prove that the perturbation
series for $W_\lambda(r,v,\lambda^{-2}t)$ converges to a limit term
by term. We describe their limit in Section \ref{sec.6}. The
convergence imposed in (\ref{3.10}) cannot hold for a sequence of
Wigner functions coming from a pure state. The
factorization (\ref{3.8}) is satisfied only for states which have
some support
in $(1 - P_+ -P_-)\mathfrak{F}$ and thus rules out bosons and
fermions. (\ref{3.8}) is a property characteristic for
boltzmannions.

For bosons and fermions one can switch freely between the BBGKY
hierarchy and the Duhamel expansion of Section \ref{sec.2}. Once we
assume the factorization (\ref{3.8}), the BBGKY hierarchy refers to
a larger class of states, not restricted to  $( P_+ + P_-)\mathfrak{F}$,
and the mapping to the Duhamel expansion of Section \ref{sec.2} is
lost. Thus in the work of Benedetto \textit{et al.} some of their
oscillatory integrals reappear in the Feynman diagrams of the
Duhamel expansion. But there are still other diagrams. Conversely
Benedetto \textit{et al.} have to consider oscillatory integrals
which do not correspond to any of the Feynman diagrams studied here.

%%%%%%%%%%%%%%%%%%%%%%%%%%%%%%%%%%%%%%%%%%%%%%%%%%%%%%%%%%%%%%

\section{A comparison with the weakly nonlinear\\
 Schr\"{o}dinger equation}\label{sec.4}
\setcounter{equation}{0}

For the Hamiltonian (\ref{2.3}) we regard $a(x)$ as a complex-valued
commutative field and, to distinguish, denote it by
$\psi:\mathbb{Z}^d\to \mathbb{C}$. The classical Hamiltonian
functional reads
\begin{equation}\label{4.1}
H=\sum_{x,y\in\mathbb{Z}^d}\alpha(x-y)\psi(x)^\ast
\psi(y)+\tfrac{1}{2}\lambda\sum_{x,y\in\mathbb{Z}^d}
V(x-y)|\psi(x)|^2 |\psi(y)|^2\,.
\end{equation}
We set
\begin{equation}\label{4.2}
\psi(x)=\frac{1}{\sqrt{2}}(q_x+\mathrm{i}p_x)
\end{equation}
and regard $q_x,p_x$ as canonically conjugate variables. Then the
field $\psi$ evolves in time as
\begin{equation}\label{4.3}
\frac{d}{dt}\psi(x,t)=-\mathrm{i}\sum_{y\in\mathbb{Z}^d}\alpha(x-y)\psi(y,t)
-\mathrm{i}\lambda\sum_{y\in\mathbb{Z}^d}V(x-y)|\psi(y,t)|^2
\psi(x,t)\,,
\end{equation}
which has the same form as (\ref{2.7a}). In particular, using the
interaction representation as in (\ref{2.9d}), the Duhamel expansion
for products of the form
$\prod^{n_0}_{j=1}\hat{\psi}(k_j,\sigma_j,t)$ is {\it identical} to
(\ref{2.20}) derived for  the quantum evolution. The only difference
resides in the average over the initial state. For the quantum case
one has
\begin{equation}\label{4.4}
\langle \prod^{n_0+2n}_{j=1} \hat{a}(k_j,\sigma_j)\rangle
\end{equation}
with an operator ordering inherited from the Feynman diagram under
consideration, while in the classical case one has to substitute
(\ref{4.4}) by
\begin{equation}\label{4.5}
\langle \prod^{n_0+2n}_{j=1}\hat{\psi}(k_j,\sigma_j)\rangle\,,
\end{equation}
where $\langle\cdot\rangle$ denotes the average over a suitable
initial probability measure on the $\psi$-field. Since the $\psi$-field
is commutative, the ordering is irrelevant.

Let us pursue this difference in more detail. For a quasifree state
it holds, compare with Definition 2.1 and Appendix A,
\begin{equation}\label{4.6}
\langle
\prod^{2n}_{j=1}\hat{a}(k_j,\sigma_j)\rangle=
\sum_{\pi \in \mathfrak{P}(2n)}\varepsilon(\pi)\prod^n_{j=1}\langle
\hat{a}(k_{\pi(j)},\sigma_{\pi(j)})\hat{a}(k_{\pi(n+j)},\sigma_{\pi(n+j)})\rangle\,.
\end{equation}
Here $ \mathfrak{P}(2n)$ is the set of all pairings of $2n$ elements
with labeling such that, in each factor of the product, the operator
order is the same as on the left hand side and $\varepsilon(\pi)=1$
for bosons, $\varepsilon(\pi)=\pm 1$ for fermions depending on
whether the permutation induced by the pairing is even or odd. Since
$\langle a^\ast a^\ast\rangle=0=\langle aa \rangle$, the average
vanishes whenever $\sum_{j=1}^{2n} \sigma_j \neq 0$. The classical
analogue of a quasifree state is a Gaussian measure for which the
only nonvanishing moments are of the form
\begin{equation}\label{4.7}
\langle \prod^{n}_{j=1}\hat{\psi}(k_j)^\ast
\hat{\psi}(k_{n+j})\rangle=\sum_{\pi \in \mathcal{P}(n)}
\prod^n_{j=1}\langle \hat{\psi}(k_j)^\ast
\hat{\psi}(k_{n+\pi(j)})\rangle\,
\end{equation}
with $\mathcal{P}(n)$ denoting the set of all permutations of $n$
elements. Note that (\ref{4.7}) agrees with the bosonic version of
(\ref{4.6}) except for operator ordering.

For bosons it holds
\begin{equation}\label{4.8}
\langle \hat{a}(k)^\ast
\hat{a}(k')\rangle=\delta(k-k')W(k)\,,\; \langle
\hat{a}(k')\hat{a}(k)^\ast
\rangle=\delta(k-k')(1+W(k))
\end{equation}
and for fermions
\begin{equation}\label{4.9}
\langle \hat{a}(k)^\ast
\hat{a}(k')\rangle=\delta(k-k')W(k)\,,\; \langle
\hat{a}(k')\hat{a}(k)^\ast
\rangle=\delta(k-k')(1-W(k))\,.
\end{equation}
If $W$ is smooth, then so is $1+\theta W(k)$, and for a subleading
Feynman diagram the particular operator order makes no difference.
On the other hand for a leading diagram one has to keep track of the
order and the limit will differ classically and quantum
mechanically, as it should be. In fact, the collision operator of
the Boltzmann equation is purely cubic for the commutative field
while it picks up an additional quadratic piece quantum mechanically
with a relative sign which depends on the statistics of the particles.\medskip\\
\textit{Remark 4.1 (Rayleigh-Jeans catastrophe).} If one replaces
$\mathbb{Z}^d$ by $\mathbb{R}^d$ and $\Omega\ast$ by $-\Delta$, then
(\ref{4.3}) turns into the Hartree equation
\begin{equation}\label{4.10}
\mathrm{i}\frac{\partial}{\partial
t}\psi(x,t)=-\Delta\psi(x,t)+\lambda\int_{\mathbb{R}^d} dy
V(x-y)|\psi(y,t)|^2 \psi(x,t)\,,
\end{equation}
and for a $\delta$-potential into the dispersive nonlinear
Schr\"{o}dinger equation (also called Gross-Pitaevskii equation)
\begin{equation}\label{4.11}
\mathrm{i}\frac{\partial}{\partial
t}\psi(x,t)=-\Delta\psi(x,t)+\lambda|\psi(x,t)|^2\psi(x,t)\,.
\end{equation}
At $\lambda=0$ the corresponding equilibrium measure is Gaussian,
gauge-invariant, and has the covariance
\begin{equation}\label{4.12}
\langle\hat{\psi}(k)^\ast\hat{\psi}(k')\rangle
=\delta(k-k')\big(\beta(k^2-\mu)\big)^{-1}\,,\quad \mu<0\,.
\end{equation}
For dimension $d=3$, the covariance (\ref{4.12}) is ultraviolet
divergent. This is the analogue of the classical Rayleigh-Jeans
catastrophe for the Maxwell field, in which case the covariance is
$\delta(k-k')|k|^{-1}$.\hspace*{0cm}\hfill{$\diamondsuit$}

%%%%%%%%%%%%%%%%%%%%%%%%%%%%%%%%%%%%%%%%%%%%%%%%%%%%%%%%%%%%%%%%%%

\section{The spatially homogeneous
 Boltzmann-Nord\-heim equation}\label{sec.6}
\setcounter{equation}{0}

We consider an initial state which is quasifree, gauge and
translation invariant, and thus completely characterized by its
two-point function
\begin{equation}\label{6.1}
\langle\hat{a}(k)^\ast\hat{a}(k')\rangle=
\delta(k-k')W(k)\,.
\end{equation}
By construction $W\geq 0$ and for fermions $W\leq 1$ in addition.
The dynamics preserves gauge and translation invariance. Therefore
\begin{equation}\label{6.2}
\langle\hat{a}(k,t)^\ast\hat{a}(k',t)\rangle=
\delta(k-k')W_\lambda(k,t)\,.
\end{equation}
As argued above, one expects that, for small $\lambda$,
$W_\lambda(k,t)$ will in approximation be governed by a nonlinear
transport
equation.\medskip\\
\textbf{Conjecture 5.1} \textit{Under suitable assumptions on $\omega$ and
on the covariance $W$ in (\ref{6.1}), it holds
\begin{equation}\label{6a.3}
\lim_{\lambda\to 0} W_\lambda(k,\lambda^{-2}t)=W(k,t)\,,
\end{equation}
$W(k,0)=W(k)$, and with this initial condition $W(t)$ satisfies the
Boltzmann-Nordheim equation
\begin{equation}\label{6a.4}
\frac{\partial}{\partial t}W(k,t)=\mathcal{C}\big(W(t)\big)(k)\,.
\end{equation}
Here the  collision operator is given by}
\begin{eqnarray}\label{6a.5}
&&\hspace{-53pt}\mathcal{C}(W)(k_1)= \pi \int_{(\mathbb{T}^d)^3}
dk_2 dk_3 dk_4 \delta(k_1+k_2-k_3-k_4)\delta(\omega_1
+\omega_2-\omega_3-\omega_4)\nonumber\\
&&\hspace{20pt}\times|\widehat{V}(k_2-k_3)+\theta
\widehat{V}(k_2-k_4)|^2 (\tilde{W}_1\tilde{W}_2 W_3 W_4-W_1
W_2\tilde{W}_3\tilde{W}_4)\,.\medskip
\end{eqnarray}
Since the expressions tend to become lengthy we use, here and in what
follows, the standard shorthand $\omega_j=\omega(k_j)$,
$W_j=W(k_j)$, $j=1,2,3,4$.\medskip\\
 \textit{Remark 5.2 (Further collision operators)}. Inserting the
definition $\tilde{W}=1+\theta W$, the Boltzmann-Nordheim collision operator becomes
\begin{eqnarray}\label{6a.6}
&&\hspace{-20pt}\mathcal{C}(W)(k_1)= \pi \int_{(\mathbb{T}^d)^3}
dk_2 dk_3 dk_4 \delta(k_1+k_2-k_3-k_4)\delta(\omega_1
+\omega_2-\omega_3-\omega_4)\nonumber\\
&&\hspace{60pt}\times
|\widehat{V}(k_2-k_3)+\theta\widehat{V}(k_2-k_4)|^2\big(
\theta(W_2 W_3 W_4 + W_1 W_3 W_4  \nonumber\\
&&\hspace{60pt}- W_1 W_2 W_4 - W_1 W_2 W_3) + W_3 W_4 - W_1
W_2\big)\,.
\end{eqnarray}
In case of the nonlinear Schr\"{o}dinger equation, $\hat{a}(k)$ is
replaced by the commutative field $\hat{\psi}(k)$. Then
$\langle\cdot\rangle$ is a translation and gauge invariant Gaussian
measure and $W(k)$ in (\ref{6.1}) defines its covariance. In this
case the collision operator reads
\begin{eqnarray}\label{6a.7}
&&\hspace{-36pt}\mathcal{C}_{\mathrm{NLS}}(W)(k_1)= \pi
\int_{(\mathbb{T}^d)^3} dk_2 dk_3 dk_4
\delta(k_1+k_2-k_3-k_4)\delta(\omega_1
+\omega_2-\omega_3-\omega_4)\\
&&\hspace{-25pt} \times|\widehat{V}(k_2-k_3)+\widehat{V}(k_2-k_4)|^2
(W_2 W_3 W_4 + W_1 W_3 W_4 - W_1 W_2 W_4 - W_1 W_2 W_3)\nonumber
\end{eqnarray}
and thus differs from (\ref{6a.4}) with $\theta=1$ only through the
quadratic terms.

If one imposes the initial condition (\ref{3.8}) corresponding to
boltzmannions, then the collision operator becomes
\begin{eqnarray}\label{6a.8}
&&\hspace{-23pt}\mathcal{C}_{\mathrm{CL}}(W)(k_1)= 2\pi
\int_{(\mathbb{T}^d)^3} dk_2 dk_3 dk_4
\delta(k_1+k_2-k_3-k_4)\delta(\omega_1
+\omega_2-\omega_3-\omega_4).\nonumber\\
&&\hspace{52pt}\times |\widehat{V}(k_2-k_3)|^2 (W_3 W_4 - W_1 W_2)\,.
\,
\end{eqnarray}
 In their set-up Benedetto \textit{et al.} prove (\ref{6a.8}) in the
sense that the perturbation series generated by the BBGKY hierarchy
(\ref{3.6}) converges term by term to the perturbation series
generated by (\ref{6a.4}) with collision operator
$\mathcal{C}_\mathrm{CL}$. Since they do not renormalize the
dispersion as in (\ref{5.11}) below, they have to impose that
$\widehat{V}(0)=0$. One recognizes (\ref{6a.8}) as the classical
Boltzmann equation with the Born approximation to the differential
cross section. Thus on the kinetic level boltzmannions behave like
classical point particles.

Note that for bosons, $\theta = 1$, 
\begin{equation}\label{6a.7a}
\mathcal{C} = \mathcal{C}_{\mathrm{NLS}} + \mathcal{C}_{\mathrm{CL}},
\end{equation}
to say, adding the collision operators for classical waves and classical particles yields the quantum mechanical collision operator.
It is surprising that quantizing either the nonlinear
Schr\"{o}dinger equation or classical point particles results in such a small modification on the level of the kinetic equation.
\hspace*{0cm}\hfill{$\diamondsuit$}\medskip\\
\textit{Remark 5.3 (Spatially inhomogeneous Boltzmann equation)}.
If one adds in (\ref{6a.4}), (\ref{6a.5}) the spatial variation,
then the Boltzmann-Nordheim equation becomes
\begin{equation}\label{6a.11}
 \frac{\partial}{\partial t}W(r,k,t)+\frac{1}{2\pi}\nabla_k\omega(k)\cdot\nabla_r
 W(r,k,t)=\mathcal{C}\big(W(r,\cdot,t)\big)(k)\,,
\end{equation}
where our notation is supposed to indicate that $\mathcal{C}$ acts
on the argument $k$ at fixed $r,t$. Of course, the same holds for
$\mathcal{C}$ replaced by $\mathcal{C}_\mathrm{NLS}$ or
$\mathcal{C}_\mathrm{CL}$. (\ref{6a.11}) can be interpreted as coming
from the motion of classical particles with kinetic energy
$\omega(k)$. Collisions between particles are implicitly defined
through the conservation of energy and momentum,
\begin{equation}\label{6a.12}
\omega_1+\omega_2=\omega_3+\omega_4\,,\quad k_1+k_2=k_3+k_4\,.
\end{equation}
The collision rule thus depends on the particular form of $\omega$
and can be very counterintuitive when viewed from the perspective of
potential scattering of mechanical particles. In the case of the
special dispersion relation $\omega(k)=k^2/2$ on $\mathbb{R}^3$,
energy and momentum conservation can be parameterized in the form
\begin{equation}\label{6a.13}
k_3=k_1-\hat{\omega}\cdot (k_1-k_2)\hat{\omega}\,,\quad
k_4=k_2+\hat{\omega}\cdot (k_1-k_2)\hat{\omega} \,,
\end{equation}
with $|\hat{\omega}|=1$, i.e., $\hat{\omega}\in S^2$. Then the
collision operator acquires the more conventional form
\begin{eqnarray}\label{6a.14}
&&\hspace{-50pt}\mathcal{C}(W)(k_1)= \int_{\mathbb{R}^3}dk_2
\int_{S^2} d\hat{\omega}|\hat{\omega}\cdot(k_1-k_2)||\widehat{V}
(\hat{\omega}\cdot(k_1-k_2)\hat{\omega})
\nonumber\\
&&\hspace{3pt}+\theta\widehat{V}(k_1-k_2-
\hat{\omega}\cdot(k_1-k_2)\hat{\omega})|^2 \big(\tilde{W}_1
\tilde{W_2}W_3 W_4- W_1 W_2 \tilde{W}_3\tilde{W}_4\big)\,.
\end{eqnarray}
For a rotational symmetric potential, $\widehat{V}(k)=\widehat{V}_\mathrm{r}(|k|)$,
the collision cross section simplifies to
\begin{equation}\label{6a.15}
|\hat{\omega}\cdot(k_1-k_2)||\widehat{V}_\mathrm{r}
(|\hat{\omega}\cdot(k_1-k_2)|)+\theta\widehat{V}_r\big(((k_1-k_2)^2-
(\hat{\omega}\cdot(k_1-k_2))^2)^{1/2}\big)|^2\,.
\end{equation}
One notes that for a smooth potential the decay is exponential and
even for hard spheres the decay is proportional to $|k_1-k_2|^{-3}$.
Thus at high energies there is only little scattering.
\hspace*{0cm}\hfill{$\diamondsuit$}\medskip\\
\textit{Remark 5.4 (History)}. Equations (\ref{6a.4}), (\ref{6a.5})
were first written down by Nordheim \cite{N29}, where he had in mind
the true quantum mechanical scattering cross section, rather than
only its Born approximation. In 1929 Peierls \cite{P29} studied
lattice vibrations with small nonlinearity both classically and
quantized. For this particular weakly nonlinear wave equation he
derives (\ref{6a.4}) with the analogue of the collision term
(\ref{6a.7}). Later on it was realized that Peierls' ideas apply to
a more general class of weakly nonlinear wave equations, e.g. see
\cite{Za}. For quantized lattice vibrations Peierls uses Fermi's
golden rule and arrives at (\ref{6a.4}) with the analogue of the
collision operator (\ref{6a.5}) for $\theta=1$. The terminology is
not uniform. In kinetic theory the name Uehling-Uhlenbeck seems to
be most frequent because they studied the equation in their
pioneering work \cite{UU33}. For phonon transport Peierls or
Boltzmann-Peierls is used. For dilute Bose gases Boltzmann-Nordheim
seems to be rather established. We follow this latter convention for
reasons of priority.\medskip\hspace*{0cm}\hfill{$\diamondsuit$}

To approach the Conjecture one expands $\langle \hat{a}(k,t)^\ast
\hat{a}(k',t)\rangle$ in the Duhamel series (\ref{2.20}) up to some
$N$ depending suitably on $\lambda$. Currently there seems to be no
good idea of how to control the error term. This leaves one with the
program of the term by term convergence, which to some extent will
be explained in Section \ref{sec.5}. An important improvement as
regards to Section \ref{sec.2} is to renormalize the bare dispersion
$\omega$ to
\begin{equation}\label{6a.9}
\omega^\lambda(k_1,t)=\omega(k_1)+\lambda\int_{\mathbb{T}^d} dk_2
W_\lambda
(k_2,t)\big(\widehat{V}(0)+\theta\widehat{V}(k_1-k_2)\big)\,.
\end{equation}
In contrast to Section \ref{sec.5}, in the present context
$\omega^\lambda$ is time-dependent through the itself unknown
$W_\lambda(t)$.

By mass conservation
\begin{equation}\label{6b.9}
\int_{\mathbb{T}^d} dk_1 W(k_1,t)=\int_{\mathbb{T}^d} dk_1
W(k_1,0)\,.
\end{equation}
Thus the term proportional to $\widehat{V}(0)$ is in fact constant.
For the second term one has the trivial estimate
\begin{equation}\label{6a.10}
\Big|\lambda\int_{\mathbb{T}^d} dk_2 W_\lambda
(k_2,t)\theta\widehat{V}(k_1-k_2)\Big|\leq
\lambda\int_{\mathbb{T}^d} dk_1 W(k_1,0)\sum_{x\in\mathbb{Z}^d}|V(x)|\,.
\end{equation}
These observations leave us with two choices: If
$\widehat{V}(k)=\widehat{V}(0)$, then the fermions become
noninteracting and one is left with lattice bosons interacting
through a quartic on-site potential. $\omega$ is renormalized by a
constant proportional to $\lambda$ which is easily taken care off.
If $\widehat{V}(k)$ depends on $k$, the situation is more
complicated. Because of the convolution, $\omega^\lambda$ depends
smoothly on $k_1$, but it could be rapidly oscillating in $t$.
Whether the bound (\ref{6a.10}) suffices to ensure the $\ell_3$ dispersivity,
the constructive interference, and the crossing bound of Section
\ref{sec.5} remains to be investigated.

In this section we discuss a more modest step, namely the leading
part of the main term. For this purpose we make the following
definitions.\medskip\\
\textbf{Definition 5.5} (pairing property). \textit{Let us consider
a Feynman diagram of even order, $n$ even, where the integration
over all momentum $\delta$-functions has been carried out. The
Feynman diagram satisfies the pairing rule if for every even time
slice the momenta are paired, i.e., to each line segment with
momentum $k$ and parity $\sigma$ there exists, in the same even time
slice another line segment with momentum $-k$ and parity
$-\sigma$.}\medskip\\
\textbf{Definition 5.6} (leading diagrams). \textit{A Feynman diagram is
\textit{leading}, if it satisfies the pairing property and if it
does not contain the factor $\widehat{V}(0)$. }\medskip\\
Note that in the 0-th
time slice $[0,s_0]$ the pairing is induced by the initial
state. Thus the structure of a leading diagram can be obtained by
iteration.

\begin{figure}[t]
\setlength{\unitlength}{1cm}
\begin{picture}
(10,7)(-1,-1.5)

\linethickness{0.1pt}
\put(0.5,0){\line(1,0){12}}
\put(0.5,1){\line(1,0){12}}
\put(0.5,2){\line(1,0){12}}
\put(0.5,3){\line(1,0){12}}
\put(0.5,4){\line(1,0){12}}
\put(0.5,5){\line(1,0){12}}

\thicklines
\put(1,-0.2){\line(0,1){1.2}}
\put(2,-0.1){\line(0,1){1.1}}
\put(2,-0.4){\line(0,1){0.1}}
\put(3,-0.10){\line(0,1){0.6}}
\put(3,-0.6){\line(0,1){0.1}}
\put(4,-0.10){\line(0,1){1.1}}
\put(4,-0.8){\line(0,1){0.1}}
\put(5,-0.10){\line(0,1){0.6}}
\put(5,-1){\line(0,1){0.1}}

\put(6,-0.10){\line(0,1){1.1}}
\put(6,-0.6){\line(0,1){0.1}}
\put(7,-0.10){\line(0,1){1.1}}
\put(7,-0.8){\line(0,1){0.3}}

\put(8,-0.10){\line(0,1){1.1}}
\put(8,-0.9){\line(0,1){0.4}}
\put(8,-1.2){\line(0,1){0.1}}

\put(9,-0.10){\line(0,1){1.1}}
\put(9,-1){\line(0,1){0.5}}
\put(10,-0.2){\line(0,1){1.2}}
\put(11,-0.4){\line(0,1){1.4}}
\put(12,-1.2){\line(0,1){2.2}}

\put(1,1){\line(0,1){1}}
\put(2,1){\line(0,1){1}}
\put(4,1){\line(0,1){0.5}}
\put(6,1){\line(0,1){1}}
\put(7,1){\line(0,1){0.5}}
\put(8,1){\line(0,1){1}}
\put(9,1){\line(0,1){1}}
\put(10,1){\line(0,1){1}}
\put(11,1){\line(0,1){1}}
\put(12,1){\line(0,1){1}}

\put(1,2){\line(0,1){1}}
\put(2,2){\line(0,1){1}}
\put(6,2){\line(0,1){1}}
\put(8,2){\line(0,1){1}}
\put(9,2){\line(0,1){0.5}}
\put(10,2){\line(0,1){1}}
\put(11,2){\line(0,1){0.5}}
\put(12,2){\line(0,1){1}}

\put(1,3){\line(0,1){0.5}}
\put(2,3){\line(0,1){1}}
\put(6,3){\line(0,1){0.5}}
\put(8,3){\line(0,1){1}}
\put(10,3){\line(0,1){1}}
\put(12,3){\line(0,1){1}}

\put(2,4){\line(0,1){1}}
\put(8,4){\line(0,1){1}}
\put(10,4){\line(0,1){1}}
\put(12,4){\line(0,1){1}}

\put(3.5,1){\line(1,0){1}}
\put(4.5,2){\line(1,0){2}}
\put(9.5,3){\line(1,0){1}}
\put(1.5,4){\line(1,0){4}}

\put(1,-0.2){\line(1,0){9}}
\put(2,-0.4){\line(1,0){9}}
\put(3,-0.6){\line(1,0){3}}
\put(4,-0.8){\line(1,0){3}}
\put(5,-1){\line(1,0){4}}
\put(8,-1.2){\line(1,0){4}}

\put(4,1){\circle*{0.2}}
\put(6,2){\circle*{0.2}}
\put(10,3){\circle*{0.2}}
\put(2,4){\circle*{0.2}}

\put(3,0.3){\vector(0,-1){0.}}
\put(4,0.3){\vector(0,-1){0.}}
\put(5,0.6){\vector(0,1){0.}}
\put(6,1.6){\vector(0,1){0.}}
\put(7,1.6){\vector(0,1){0.}}
\put(8,0.6){\vector(0,1){0.}}
\put(9,2.3){\vector(0,-1){0.}}
\put(10,2.6){\vector(0,1){0.}}
\put(11,2.6){\vector(0,1){0.}}
\put(12,0.3){\vector(0,-1){0.}}
\put(4,1.3){\vector(0,-1){0.}}

\put(1,2.3){\vector(0,-1){0.}}
\put(2,2.3){\vector(0,-1){0.}}
\put(6,2.6){\vector(0,1){0.}}
\put(9,2.3){\vector(0,-1){0.}}
\put(10,2.6){\vector(0,1){0.}}
\put(11,2.6){\vector(0,1){0.}}

\put(2,4.3){\vector(0,-1){0.}}
\put(8,4.6){\vector(0,1){0.}}
\put(10,4.6){\vector(0,1){0.}}
\put(12,4.3){\vector(0,-1){0.}}

\put(6.6,-0.2){\vector(1,0){0.}}
\put(7.6,-0.4){\vector(1,0){0.}}
\put(4.6,-0.6){\vector(1,0){0.}}
\put(5.6,-0.8){\vector(1,0){0.}}
\put(7.4,-1.0){\vector(-1,0){0.}}
\put(10.4,-1.2){\vector(-1,0){0.}}

\put(3.5,0.5){\oval(1,1)[tl]}
\put(4.5,0.5){\oval(1,1)[tr]}
\put(4.5,1.5){\oval(1,1)[tl]}
\put(6.5,1.5){\oval(1,1)[tr]}
\put(9.5,2.5){\oval(1,1)[tl]}
\put(10.5,2.5){\oval(1,1)[tr]}
\put(1.5,3.5){\oval(1,1)[tl]}
\put(5.5,3.5){\oval(1,1)[tr]}

\end{picture}
\caption{A leading Feynman diagram with $n_0 = 4$, $n = 4$, and
interaction history (3,3,5,1). Up (down) arrows correspond to parity
$+1$ ($-1$). The pairing from the initial state is indicated, where
right (left) pointing arrow stands for the order $\langle a^\ast
a\rangle$ ($\langle a a^\ast\rangle$).}\label{fig2}

\end{figure}
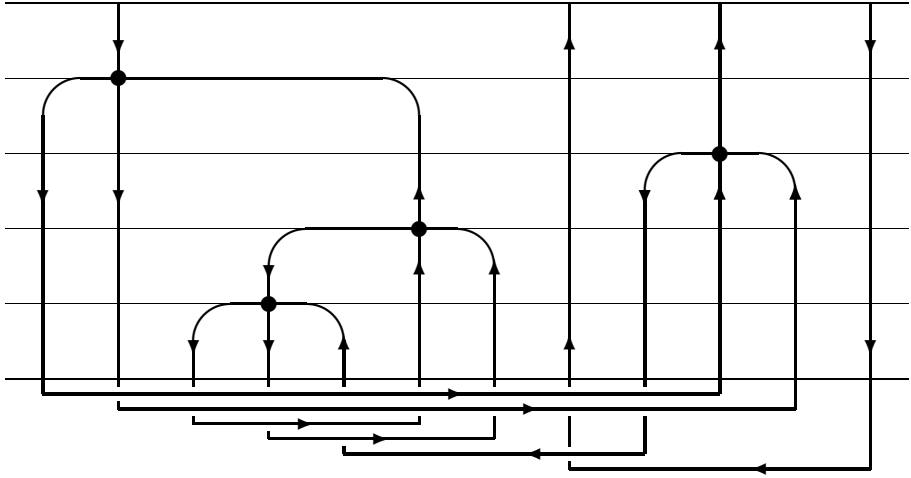

So let us assume that for the time slice with even label $i$ all
line segments are paired. If there are $2n$ line segments, they
carry momenta $\sigma_j k_j$, $j=1,\ldots,n$, $\sigma_j=\pm 1$.
$k_j$ and $-k_j$ are paired. At the end of time slice $i$ three
neighboring lines fuse to the line segment with momentum $\sigma'
k'$. If two momenta would be paired, the diagram necessarily contains
the factor $\widehat{V}(0)$. Thus we label neighboring lines by three
distinct momenta  $\sigma_j k_j$,
$j=1,2,3$. In the next step there can be two cases which are
both displayed in Fig.\ \ref{fig2}. Case a) corresponds to the
$3^\textrm{rd}$ and $4^\textrm{th}$ fusion, while case b)
corresponds to the $1^\textrm{st}$ and $2^\textrm{nd}$ fusion. In
case a) the line segment $\sigma' k'$ does not participate in the
fusion at the end of time slice $i+1$. The fusing triplet has
momenta $\sigma_{\pi(j)}k_{\pi(j)}$, $j=1,2,3,$ and the momentum of
the fused line segment is $\sigma'' k''$. By the same argument as above, the
$\pi(j)$'s must be distinct. To have pairing in time slice $i+2$
requires $k'=k''$, $\sigma'=-\sigma''$. In turn this is possible
only if $\{1,2,3\}=\{\pi(1),\pi(2),\pi(3)\}$. In case b) the line
segment $\sigma'k'$ participates in the fusion at the end of time
slice $i+1$. The fusing momenta are $\sigma'k'$, $\sigma_{\pi(1)}
k_{\pi(1)}$, $\sigma_{\pi(2)} k_{\pi(2)}$. The fused momentum is
$\sigma''k''$. In slice $i+2$, $\sigma''k''$ must be paired with,
say, $\sigma_{\pi(3)} k_{\pi(3)}$. There will be a factor
$\widehat{V}(0)$ unless all three $\pi(j)$'s are distinct. The
pairing of $\sigma''k''$ and $\sigma_{\pi(3)} k_{\pi(3)}$ is
possible only if $\{1,2,3\}=\{\pi(1),\pi(2),\pi(3)\}$.

\begin{figure}[t]

\setlength{\unitlength}{1cm}
\begin{picture}
(10,4.5)(-3.5,-1)

\linethickness{0.1pt}
\put(0.5,0){\line(1,0){6}}
\put(0.5,1){\line(1,0){6}}
\put(0.5,2){\line(1,0){6}}
\put(0.5,3){\line(1,0){6}}

\thicklines
\put(1,0){\line(0,1){3}}
\put(2,0){\line(0,1){1.5}}
\put(3,0){\line(0,1){0.5}}
\put(4,0){\line(0,1){0.5}}
\put(5,0){\line(0,1){1.5}}
\put(6,0){\line(0,1){3}}

\put(2,1){\line(1,0){1.5}}
\put(1,2){\line(1,0){3.5}}

\put(1.5,1.5){\oval(1,1)[tr]}
\put(4.5,1.5){\oval(1,1)[tr]}
\put(2.5,0.5){\oval(1,1)[tr]}
\put(3.5,0.5){\oval(1,1)[tr]}

\put(0.8,-0.5){$+$}
\put(1.8,-0.5){$-$}
\put(2.8,-0.5){$+$}
\put(3.8,-0.5){$+$}
\put(4.8,-0.5){$+$}
\put(5.8,-0.5){$-$}
\put(1.2,2.5){$+$}
\put(5.5,2.5){$-$}
\put(1.5,1.3){$-$}

\put(1,2){\circle*{0.2}}
\put(2,1){\circle*{0.2}}

\end{picture}
\caption{Contraction of the diagram from Figure 2. The order parity,
$\tau$, of each contracted line is indicated.}\label{fig3}
\end{figure}
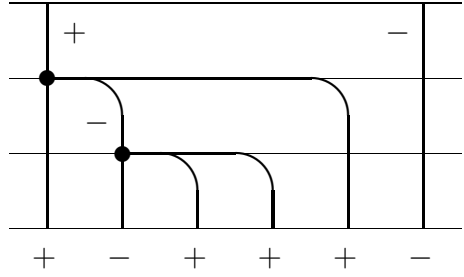

We conclude that for a leading Feynman diagram there must be three
paired line segments in slice $i$ which connect through two fusions
to a single paired line segment in slice $i+2$, $i=0,2,\ldots,n$,
$n$ even, compare with Fig.\ \ref{fig2}. This property allows us to
represent a leading diagram through a contracted diagram which we
explain next, see Fig.\ \ref{fig3}.

In a contracted diagram we draw only the even time slices and each
pair as a single line segment. A line segment thus carries a
momentum $k$. But we still have to distinguish the relative order
within the pair. If in the original Feynman diagram the order from left
to right is $-,+$, then the line carries the order parity $\tau=+
1$, while for the order $+,-$ the order parity is $\tau=-1$. In the
contracted diagram, at the end of each time slice a triplet of
neighboring line segments fuses into a single line segment. In our
case, if the leading Feynman diagram has order $2n$, then each
contracted diagram has $n$ collisions and there are $(2n+1)!/2^nn!$
contracted diagrams.

To compute the vertex strength for the contracted diagram one has to
sum over all Feynman diagrams which fuse 3 pairs at even time slice
into the single pair at time slice $i+2$. There are 16 such
diagrams. Their sum yields the vertex strength given below in
(\ref{6.3}).

To write down the integral corresponding to a contracted diagram, it
is useful to introduce the ``correlation'' functions
$\widehat{\rho}_n:(\mathbb{T}^d\times\{-1,1\})^n\to\mathbb{C}$ and
to define time-dependent collision operators
$\mathcal{C}_{j,n+2}(t)$ with $j=1,\ldots,n$. We set
\begin{eqnarray}\label{6.3}
&&\hspace{-20pt}\big(\mathcal{C}_{j,n+2}(t)\widehat{\rho}_{n+2}\big)
(k_1,\tau_1,\ldots,k_n,\tau_n)=2\int_{(\mathbb{T}^d)^3}
dg_2 dg_3 dg_4 \delta(k_j+g_2-g_3-g_4)\nonumber\\
&&\hspace{-10pt}\times\cos\big((\omega(k_j)+\omega(g_2)-\omega(g_3)-\omega(g_4))t\big)
\widehat{V}(g_2-g_3)\big(\widehat{V}(g_2-g_3)+\theta\widehat{V}(g_2-g_4)\big)\nonumber\\
&&\hspace{-10pt}\times\big(\widehat{\rho}_{n+2}(\ldots,
g_2,-\tau_j,\ldots,g_3,\tau_j,g_4,\tau_j)
+\theta\widehat{\rho}_{n+2}(\ldots,k_j,\tau_j,\ldots,g_3,1,g_4,1)\nonumber\\
&&\hspace{-0pt}-\widehat{\rho}_{n+2}(\ldots,
k_j,\tau_j,\ldots,g_2,1,g_3,-1) -\theta\widehat{\rho}_{n+2}(\ldots,
k_j,\tau_j,\ldots,g_2,1,g_4,1)\big)\,,
\end{eqnarray}
which acts on the arguments $j,n+1,n+2$ of $\widehat{\rho}_{n+2}$,
and
\begin{equation}\label{6.4}
\mathcal{C}_{n+2}(t)=\sum^n_{j=1}\mathcal{C}_{j,n+2}(t)\,.
\end{equation}

Let
\begin{equation}\label{6.4a}
\mathcal{M}^\lambda_n(k_1,t) \delta(k_0-k_1)
\end{equation}
be the sum over all contracted diagrams at order $n$ with time span
$t$. Then
\begin{eqnarray}\label{6.4b}
&&\hspace{-22pt}\mathcal{M}^\lambda_n(k_1,\lambda^{-2}t)\nonumber\\
&&\hspace{-17pt}=\lambda^{2n}\int_{(\mathbb{R}_+)^{n+1}}d\underline{t}
\int_{(\mathbb{R}_+)^n}d\underline{s}\delta(\sum^{n+1}_{j=1}
t_j+\sum^n_{j=1} s_j-\lambda^{-2}t)\big( \mathcal{C}_3(s_1)\ldots
\mathcal{C}_{2n+1}(s_n)\widehat{\rho}_{2n+1}\big)(k_1,1)\nonumber\\
&&\hspace{-17pt}=\int_{0\leq\sum^n_{j=1}t_j\leq t}d\underline{t}
\int_{0\leq\sum^n_{j=1}s_j\leq
\lambda^{-2}(t-\sum^n_{j=1}t_j)}d\underline{s}\big(
\mathcal{C}_3(s_1)\ldots
\mathcal{C}_{2n+1}(s_n)\widehat{\rho}_{2n+1}\big)(k_1,1)\,.
\end{eqnarray}
In our particular case the input function is given by
\begin{equation}\label{6.5}
\widehat{\rho}_n(k_1,\tau_1,\ldots,k_n,\tau_n)= \prod^n_{j=1}
W(k_j,\tau_j)
\end{equation}
with
\begin{equation}\label{6.6}
W(k,1)=W(k)\,,\quad W(k,-1)=1+\theta W(k) = \tilde{W}(k)\,.
\end{equation}

To prove the limit $\lambda\to 0$ one needs a first assumption on
$\omega$.\medskip\\
\textbf{Assumption A1} ($\ell_3$ dispersivity). \textit{Let us define
\begin{equation}\label{6.6a}
p_t(x)=\int_{\mathbb{T}^d}dk
\mathrm{e}^{-\mathrm{i}t\omega(k)}\mathrm{e}^{\mathrm{i}2\pi x\cdot
k}\,.
\end{equation}
Then there exist $c > 0$, $\delta>0$ such that
\begin{equation}\label{6.7}
\sum_{x\in\mathbb{Z}^d}|p_t(x)|^3\leq c\langle t \rangle^{-1-\delta}
\end{equation}
with the shorthand $\langle t \rangle = \sqrt{1 + t^2}$.}\medskip\\
As in \cite{LuSp} one proves the following bound.\medskip\\
\textbf{Proposition 5.7} \textit{Let
$\|\rho_n\|_1\leq (c_0)^n$, where $\|\cdot\|_1$ is the
$\ell_1(\mathbb{Z}^d)^{\otimes n}$ norm in position space. Then
there exists a constant $c$ such that
\begin{equation}\label{6.8}
\int^\infty_0
d\underline{s}\sup_{k_1\in \mathbb{T}^d}\big|\big(\mathcal{C}_3(s_1)\ldots
\mathcal{C}_{2n+1}(s_n)\widehat{\rho}_{2n+1}\big)(k_1,1)\big|\leq
c^n n!\,.
\end{equation}
}\medskip\\
Note that $\mathcal{C}_{2n+1}(s)$ has $2n-1$ terms of equal size.
The $n!$ in (\ref{6.8}) thus results from the product of collision
operators.

With this bound one can introduce the collision operator
\begin{eqnarray}\label{6.8a}
&&\hspace{-30pt}(\mathcal{C}\widehat{\rho}_3(k_1,\tau)= 2\pi \int_{(\mathbb{T}^d)^2}
dk_2 dk_3
\delta\big(\omega(k_1)+\omega(k_2)-\omega(k_3)-\omega(k_1+k_2-k_3)\big)\nonumber\\
&&\hspace{10pt}\times
\widehat{V}(k_2-k_3)\big(\widehat{V}(k_2-k_3)+\theta\widehat{V}(k_2-k_4)\big)
\big(\widehat{\rho}_3(k_2,-\tau,k_3,\tau,k_1+k_2-k_3,\tau)\nonumber\\
&&\hspace{15pt}+\theta\widehat{\rho}_3(k_1,\tau,k_2,1,k_1+k_2-k_3,1)-
\widehat{\rho}_3(k_1,\tau,k_2,1,k_3,-1)\nonumber\\
&&\hspace{15pt}-\theta\widehat{\rho}_3(k_1,\tau,k_2,1,k_1+k_2-k_3,1)\big)\,,
\end{eqnarray}
where the $\delta$-function is defined through the limit
\begin{equation}\label{6.9}
\delta(\Omega)=\lim_{\varepsilon\to
0}\frac{1}{\pi}\frac{\varepsilon}{\Omega^2 +\varepsilon^2}\,.
\end{equation}
As before, from $\mathcal{C}$ we construct $\mathcal{C}_{j,n+2}$,
which is $\mathcal{C}$ now acting on the variables
$k_j,k_{n+1},$ $k_{n+2}$, and
\begin{equation}\label{6.10}
\mathcal{C}_{n+2}=\sum^n_{j=1}\mathcal{C}_{j,n+2}\,.
\end{equation}

Under the conditions of Proposition 5.7 one concludes that
\begin{equation}\label{6.11}
\lim_{\lambda\to 0}\mathcal{M}^\lambda_n(k_1,\lambda^{-2}
t)=\frac{1}{n!}t^n(\mathcal{C}_3\ldots
\mathcal{C}_{2n+1}\widehat{\rho}_{2n+1})(k_1,1)
\end{equation}
uniformly in $k_1$ and
\begin{equation}\label{6.11a}
    \sup_{k_1\in \mathbb{T}^d}|(\mathcal{C}_3\ldots
\mathcal{C}_{2n+1}\widehat{\rho}_{2n+1})(k_1,1)|\leq c^n n!\,.
\end{equation}
Thus the sum over $n$ reads
\begin{equation}\label{6.16}
W(k,t)=\sum^\infty_{n=0}\frac{t^n}{n!}(\mathcal{C}_3\ldots
\mathcal{C}_{2n+1}\widehat{\rho}_{2n+1})(k,1)\,.
\end{equation}
By (\ref{6.8}) there exists then a $t_0$, such that the sum converges provided
\begin{equation}\label{6.12}
0\leq t< t_0\,.
\end{equation}
If of interest, the time $t_0$ can be computed explicitly. Up to
numerical factors of order 1, $t_0$ is proportional to
\begin{equation}\label{6.13}
\sum_{x\in\mathbb{Z}^d}|V(x)|\quad\textrm{and}\quad \int^\infty_0
dt\sum_{x\in\mathbb{Z}^d}|p_t(x)|^3\,.
\end{equation}

Because of the particular initial conditions (\ref{6.5}),
(\ref{6.6}), the limit in (\ref{6.16}) can be written more
concisely. One recognizes $W(k,t)$ as the power series solution of
the nonlinear Boltzmann-Nordheim equation (\ref{6a.4}) with
collision operator
\begin{eqnarray}\label{6.15}
&&\hspace{-53pt}\mathcal{C}(W)(k_1)= 2\pi
\int_{(\mathbb{T}^d)^3} dk_2 dk_3 dk_4
\delta(k_1+k_2-k_3-k_4)\delta(\omega_1
+\omega_2-\omega_3-\omega_4)\nonumber\\
&&\hspace{20pt} \times\widehat{V}(k_2-k_3)\big(\widehat{V}(k_2-k_3)+
\theta\widehat{V}(k_2-k_4)\big)\nonumber\\
&&\hspace{20pt}\times(\tilde{W}_2 W_3 W_4+ \theta W_1 W_3 W_4-W_1
W_2\tilde{W}_3-\theta W_1 W_2 W_4)\,,
\end{eqnarray}
which is identical to (\ref{6a.5}).

%%%%%%%%%%%%%%%%%%%%%%%%%%%%%%%%%%%%%%%%%%%%%%%%%%%%%%%%%%%%%%%%

\section{Two-point time correlations in thermal equilibrium}\label{sec.5}
\setcounter{equation}{0}

\textit{6.1 Set-up}. In this and the following section we consider a quantum fluid in
thermal equilibrium. We fix some inverse temperature $\beta>0$ and a
chemical potential $\mu\in \mathbb{R}$ such that
\begin{equation}\label{5.1}
\omega(k)-\mu>0
\end{equation}
for all $k\in\mathbb{T}^d$. $\langle\cdot\rangle_{\beta,\lambda}$ denotes
the expectation value with respect to the $\beta$-KMS state for the
hamiltonian $H-\mu N$, $N$ the number operator. The two-point
function is denoted by
\begin{eqnarray}\label{5.1a}
&&\hspace{-25pt}\langle \hat{a}(k)^\ast
\hat{a}(k')\rangle_{\beta,\lambda} = \delta(k-k')W_{\beta,\lambda}^\theta(k)\,,
\nonumber\\
&&\hspace{-25pt} \langle \hat{a}(k)
\hat{a}(k')^\ast\rangle_{\beta,\lambda} =
\delta(k-k')\tilde{W}_{\beta,\lambda}^\theta(k)=\delta(k-k')\big(1+\theta
W_{\beta,\lambda}^\theta(k)\big)\,.
\end{eqnarray}
In the limit $\lambda\to 0$, the state $\langle\cdot\rangle_{\beta,0}$ is
quasifree with
\begin{equation}\label{5.2}
\langle \hat{a}(k)^\ast \hat{a}(k')\rangle_{\beta,0} =
\delta(k-k')W_\beta^\theta(k)\,.
\end{equation}
Since $W^\theta_\beta$ is smooth, the two-point function decays
exponentially and by quasifreeness all fully truncated correlation
functions of order greater than two vanish.

In this section we study the effective propagator as the most basic 
two-point function, namely
\begin{equation}\label{5.3}
\langle \hat{a}(k)^\ast \hat{a}(k',t)\rangle_{\beta,\lambda} =
\langle \hat{a}(k)^\ast \hat{a}(k')\rangle_{\beta,\lambda}
C_\lambda(k,t)\,,
\end{equation}
which defines $C_\lambda(k,t)$ since both sides of (\ref{5.3}) are
proportional to $\delta(k-k')$. More complicated equilibrium time
correlation functions will be discussed in Section 7. Because of the
$a$-term, clearly, $C_\lambda$ will contain the oscillatory factor
$\mathrm{e}^{-\mathrm{i}\omega(k)t}$ varying on time scale 1. To
first order in the Duhamel expansion one has
\begin{eqnarray}\label{5.4}
&&\hspace{-25pt}\mathrm{e}^{\mathrm{i}\omega(k_1)t}\langle
\hat{a}(k_0)^\ast \hat{a}(k_1,t)\rangle_{\beta,\lambda} = \langle
\hat{a}(k_0)^\ast \hat{a}(k_1)\rangle_{\beta,\lambda} \nonumber\\
&&\hspace{-5pt}-\mathrm{i}\lambda\int^t_0 ds
\int_{(\mathbb{T}^d)^3}
dk_2 dk_3 dk_4 \delta(k_1+k_2-k_3-k_4)\widehat{V}(k_2-k_3)\nonumber\\
&&\hspace{7pt}\times \exp
[\mathrm{i}s(\omega_1+\omega_2-\omega_3-\omega_4)] \langle
\hat{a}(k_0)^\ast \hat{a}(k_2)^\ast \hat{a}(k_3)
\hat{a}(k_4)\rangle_{\beta,\lambda}+\mathcal{O}(\lambda^2) \,.
\end{eqnarray}
We write
\begin{eqnarray}\label{5.5}
&&\hspace{-25pt}\langle \hat{a}(k_0)^\ast \hat{a}(k_2)^\ast
\hat{a}(k_3) \hat{a}(k_4)\rangle_{\beta,\lambda}= \langle
\hat{a}(k_0)^\ast \hat{a}(k_4)\rangle_{\beta,\lambda}
\langle \hat{a}(k_2)^\ast \hat{a}(k_3)\rangle_{\beta,\lambda}\nonumber\\
&&\hspace{-25pt}+\theta \langle \hat{a}(k_0)^\ast
\hat{a}(k_3)\rangle_{\beta,\lambda} \langle \hat{a}(k_2)^\ast
\hat{a}(k_4)\rangle_{\beta,\lambda}+ \langle \hat{a}(k_0)^\ast
\hat{a}(k_2)^\ast \hat{a}(k_3)
\hat{a}(k_4)\rangle^T_{\beta,\lambda} \,.
\end{eqnarray}
The truncated part is $\mathcal{O}(\lambda)$ and, when first integrated 
over momenta as
in (\ref{5.4}), is absolutely integrable in $s$. Therefore this contribution will be
part of the error term. The first two terms inserted in (\ref{5.4})
yield $-\mathrm{i}\lambda t \langle \hat{a}(k_0)^\ast
\hat{a}(k_1)\rangle_{\beta,\lambda}R_\lambda(k_1)$, where
\begin{equation}\label{5.6}
R_\lambda(k_1) = \int_{\mathbb{T}^d} dk_2
W_\lambda^\theta
(k_2)\big(\widehat{V}(0)+\theta\widehat{V}(k_1-k_2)\big)\,.
\end{equation}

Our computation suggests that $C_\lambda$ has a second oscillatory
factor of the form
\begin{equation}\label{5.8}
\mathrm{e}^{-\mathrm{i}\lambda R_\lambda(k)t}
\end{equation}
varying on the time scale $\lambda^{-1}$. The kinetic time scale is
order $\lambda^{-2}$. Thus it suffices to expand up to $\lambda$ as
\begin{equation}\label{5.9}
R_\lambda(k_1)= R_0(k_1)+\lambda R_1(k_1)+\mathcal{O}(\lambda^2)
\end{equation}
with
\begin{eqnarray}\label{5.10}
&&\hspace{-25pt}R_0(k_1)=\int_{\mathbb{T}^d} dk_2
W_\beta^\theta(k_2)\big(\widehat{V}(0)+\theta\widehat{V}(k_1-k_2)\big)\,,\nonumber\\
&&\hspace{-25pt} R_1(k_1)=-\beta\int_{\mathbb{T}^d} dk_2
W_\beta^\theta(k_2)\tilde{W}_\beta^\theta(k_2)
\big(\widehat{V}(0)+\theta\widehat{V}(k_1-k_2)\big)\nonumber\\
&&\hspace{32pt} \times\int_{\mathbb{T}^d} dk_3 W_\beta^\theta(k_3)
\big(\widehat{V}(0)+\theta\widehat{V}(k_1-k_3)\big)\,.\nonumber\\
\end{eqnarray}

Potentially the oscillatory term (\ref{5.8}) could be dangerous
because in an expansion in $\lambda$ it may mask the kinetic terms.
As we will see, fortunately, $R_\lambda$ can be absorbed by
renormalizing $\omega$ to
\begin{equation}\label{5.11}
\omega^\lambda=\omega+\lambda R_\lambda\,.
\end{equation}

The second order term of the Duhamel expansion will not be written
out explicitly. It consists of 18 diagrams, 12 of which combine to
\begin{equation}\label{5.12}
-\tfrac{1}{2}\lambda^2 t^2 R_0(k)^2\,.
\end{equation}
The remaining 6 diagrams sum up to
\begin{equation}\label{5.13}
-\delta(k-k')W_\beta^\theta(k)\nu(k)\lambda^2 t
\end{equation}
valid for large $t$. The decay coefficient $\nu$ is obtained as
\begin{eqnarray}\label{5.14}
&&\hspace{-20pt}\nu(k_1)=-\int^\infty_0 dt
\int_{(\mathbb{T}^d)^3}dk_2 dk_3 dk_4
\delta(k_1+k_2-k_3-k_4)\exp[\mathrm{i}t(\omega_1+\omega_2-\omega_3-\omega_4)]
\nonumber\\
&&\hspace{-10pt}\times\widehat{V}(k_2-k_3)
\big(\widehat{V}(k_2-k_4)+
\theta\widehat{V}(k_2-k_3)\big)(W^{\theta}_{\beta,3}W^{\theta}_{\beta,4}-
W^{\theta}_{\beta,2}W^{\theta}_{\beta,4}-\theta W^{\theta}_{\beta,2}
\tilde{W}^{\theta}_{\beta,3})\nonumber\\
\end{eqnarray}
with the shorthand $W^{\theta}_{\beta, j}=W_\beta^\theta(k_j)$. For the real part
of $\nu$ one obtains
\begin{eqnarray}\label{5.15}
&&\hspace{-25pt}\Re\nu(k_1)=\pi \int_{(\mathbb{T}^d)^3}dk_2 dk_3
dk_4 \delta(k_1+k_2-k_3-k_4)
\delta(\omega_1+\omega_2-\omega_3-\omega_4)\nonumber\\
&&\hspace{28pt} \times\tfrac{1}{2}\big(\widehat{V}(k_2-k_3)+ \theta
\widehat{V}(k_2-k_4)\big)^2(W^{\theta}_{\beta,1})^{-1}\tilde{W}^{\theta}_{\beta,2}
W^{\theta}_{\beta,3}W^{\theta}_{\beta,4}\,.
\end{eqnarray}
In particular
\begin{equation}\label{5.16}
\Re \nu (k_1)> 0\,.
\end{equation}

On the basis of this second order expansion the obvious conjecture
is that
\begin{equation}\label{5.17}
C_\lambda(k,t)\cong
\exp\big[-(\mathrm{i}\omega^\lambda+\lambda^2\nu(k))t\big]
\end{equation}
for $t\geq 0$ and $t=\mathcal{O}(\lambda^{-2})$, the case $t\leq 0$
following from time reversal as
\begin{equation}\label{5.18}
C_\lambda(k,t)^\ast=C_\lambda(k,-t)\,.
\end{equation}

Our goal is to prove (\ref{5.17}). For this purpose we have to
assume cluster properties of the equilibrium state and the decay of
certain oscillatory integrals.\medskip\\
\textit{6.2  $\ell_1$-clustering of fully truncated correlation
functions.} For the KMS state $\langle\cdot\rangle_{\beta,\lambda}$ we
consider the fully truncated correlation functions denoted by
$\langle\prod^n_{j=1} a(x_j,\sigma_j)\rangle^T_{\beta,\lambda}$. We refer to
Appendix A for their definition. They vanish whenever
$\sum^n_{j=1}\sigma_j\neq 0$ and, as proved in Appendix A, 
for $n\geq 4$ they do not depend on
the operator ordering except for an overall sign.\medskip\\
\textbf{Assumption A2} ($\ell_1$-clustering). \textit{Let
$\beta>0$ and $\mu$ satisfy (\ref{5.1}). There exists $\lambda_0>0$
and $c_0>0$ independent of $n$ such that for $0<\lambda\leq
\lambda_0$. and all $n\geq 4$ one has the bound
\begin{equation}\label{5.19}
\sum_{x\in(\mathbb{Z}^d)^n} \delta_{x_1 0}\Big|\langle \prod^n_{j=1}
a(x_j,\sigma_j)\rangle^T_{\beta,\lambda}\Big|\leq \lambda(c_0)^n n!\,.
\end{equation}
In addition,}
\begin{equation}\label{5.20}
\sum_{x\in\mathbb{Z}^d} |\langle a(0)^\ast{a}(x)
\rangle_{\beta,\lambda} - \langle a(0)^\ast{a}(x)
\rangle_{\beta}| \le \lambda 2 (c_0)^2 \,.\medskip
\end{equation}

Ginibre \cite{Gi} studies $\ell_1$-clustering for low density
quantum gases in the continuum, i.e., for position space $\mathbb{R}^3$ and
dispersion relation $\omega(k)=k^2$. It is rather likely that his analysis could be
carried through also for lattice gases. The $L^1$-bound by Ginibre
is based on an expansion with respect to the fugacity, hence the
small coupling regime is not optimally covered. In particular, the
prefactor $\lambda$ in (\ref{5.19}) cannot be deduced
from \cite{Gi}. It seems that extra work is needed.\medskip\\
\textit{6.3 Oscillatory integrals.} The $\ell_3$-dispersivity has been stated
already in Assumption A1. In addition we need an assumption which
controls the constructive
interference between two frequencies.\medskip\\
\textbf{Assumption A3} (constructive interference). \textit{
There exists a set $\Msing \subset \T^d$ consisting
of a union of a finite number of closed, one-dimensional, smooth submanifolds,
and a constant $C$ such that
for all $t\in \mathbb{R}$, $k_0\in \T^d$, and $\sigma\in \{\pm 1\}$,
\begin{equation}\label{5.20a}
  \Bigl| \int_{\T^d}\! d k\, \rme^{-\ci  t (\omega(k)+\sigma
    \omega(k-k_0))}\Bigr| \le \frac{C\sabs{t}^{-1} }{d(k_0,\Msing)}\, ,
\end{equation}
where $d(k_0,\Msing)$ is the distance  of $k_0$ from $\Msing$. 
}\medskip\\
\textit{Remark 6.1 (Dimension)}. Assumption A3 allows us to cut
out a small tube around each curve. If one would cut out too much,
e.g. two-dimensional surfaces, this will show up in other parts of
the proof. For this reason Assumption A3 as stated requires in
addition $d\geq 4$. If (\ref{5.20a}) would hold for $\Msing$ merely a
collection of a finite number of points, then we could accommodate
$d\geq 3$.
\medskip\hspace*{0cm}\hfill{$\diamondsuit$}

Next we need a mechanism which allows to distinguish between leading
and subleading diagrams. For the linear Schr\"{o}dinger equation
with a random potential, this mechanism has been identified in
\cite{EY00}. In the graphical representation developed in this
article, it corresponds to the crossing of two edges. 
Our condition is the natural 
generalization of the crossing estimate in \cite{EY00}.\medskip\\
\textbf{Assumption A4} (crossing bounds). \textit{
Define for $t_0,t_1,t_2\in \mathbb{R}$, $u_1,u_2\in \T^d$, and $x\in \mathbb{Z}^d$,
\begin{equation}\label{eq:defp2tx}
    K(x;t_0,t_1,t_2,u_1,u_2) = 
    \int_{\T^d} \! d k\, \rme^{\ci 2\pi x\cdot k} 
    \rme^{-\ci \big(t_0 \omega(k)+t_1 \omega(k+u_1)+ t_2 \omega(k+u_2)\big)} \, .
\end{equation}
We assume that there is 
a measurable function $\Fbcr:\T^d \times \mathbb{R}_+ \to [0,\infty]$
so that constants  $0<\gamma\le 1$, $c_1,c_2$
for the following bounds can be found.\smallskip\\
(i) 
 For any $u_i\in \T^d$, $\sigma_i\in \{\pm 1\}$, $i=1,2,3$,
and  $0<\beta\le 1$, the following bounds are satisfied:
%are bounded by   $\beta^{\gamma-1}  \Fbcr(u;\beta)$:
\begin{eqnarray}
&&\hspace{-20pt}   
 \int_{\mathbb{R}^2} d s_1 d s_2\, \rme^{-\beta (|s_1|+|s_2|)}
 \| K(s_1+s_2,\sigma_1 s_2,\sigma_2 s_2,u_1,u_2)\|_3  \|p_{s_1+s_2}\|_3^2 
 \nonumber \\ 
&&\hspace{20pt} 
 \le \beta^{\gamma-1}  \Fbcr(u_2-u_1;\beta) \, ,
\label{eq:crossingest1a}
\\  &&\hspace{-20pt}
 \int_{\mathbb{R}^2} d s_1 d s_2\, \rme^{-\beta (|s_1|+|s_2|)}
 \prod_{i=1}^3 \|K(s_1+s_2,\sigma_i s_2,0,u_i,0)\|_3 
 \nonumber \\ 
&&\hspace{20pt}
\le \beta^{\gamma-1}  \Fbcr(u_n;\beta),\quad \text{for any }n\in\{1,2,3\}\, .
\label{eq:crossingest1b} 
\end{eqnarray}
(ii)
 For all   $0<\zeta\le \zeta_0$ we have
\begin{equation}\label{eq:crossingest2a}
 \int_{\T^d} d k\, \Fbcr(k;\zeta) \le c_1 \sabs{\ln \zeta}^{c_2}  ,
\end{equation}
and if also $u,k_0\in \T^d$, 
  $\alpha\in \R$, 
  $\sigma\in\{\pm 1\}$, and $n\in \{1,2,3\}$,
and we denote  $k=(k_1,k_2,k_0-k_1-k_2)$, then
\begin{equation}\label{eq:crossingest2}
  \int_{(\T^d)^2} d k_1d k_2\, \Fbcr(k_n+u;\zeta)
\frac{1}{|\alpha-\Omega(k,\sigma)+\ci\zeta|} 
%\nonumber \\ & \qquad\times
 \le c_1 \sabs{\ln \zeta}^{1+c_2}  ,
\end{equation}
where $\Omega:(\T^d)^3\times \{\pm 1\}\to \R$ is defined by}
\begin{equation}\label{eq:defOmega} 
 \Omega(k,\sigma) 
= \omega(k_3)-\omega(k_1)+\sigma(\omega(k_2)-\omega(k_1+k_2+k_3))\, .
\end{equation}
\textit{Remark 6.2 (Reduction)}. The reader may wonder whether the
stated assumptions can be proved for a specified class of
$\omega$'s. Let us first emphasize that the reduction of the many,
very high-dimensional oscillatory integrals to a few low dimensional
oscillatory integrals involving only $\omega$ is already a big step
in the right direction. It is also clear that the reduction cannot
be pushed any further. The $\ell_3$ dispersivity is needed already to define
the limit equation. Constructive interference occurs always at
$k_0=0$. Thus some control of the phenomenon is required. The
crossing bound reflects the mechanism how only a few diagrams
survive in the limit. At present, A4 is one version which works, but its 
optimal form could eventually look differently. In any case, Assumptions A1, A3, and A4
rely on results from a disjoint mathematical discipline. In this
sense, the situation is similar to A2. Clustering is proved within
rigorous statistical mechanics. In our context it is a necessary
input stated in a form which is regarded as obvious by the experts.
A1 can be proved by stationary phase methods and holds generically
for dimension $d\geq 3$. We are not aware that A3 has ever been
studied systematically, but hope that our work might serve as a
motivation. A4 in the somewhat simpler context of the random
Schr\"{o}dinger equation has been proved for nearest neighbor
couplings in \cite{Ch05} and is investigated more systematically in
\cite{Lu06,ES07}.\medskip\hspace*{0cm}\hfill{$\diamondsuit$}\\
\textit{Remark 6.3 (Example)}. For on-site and nearest neighbor 
couplings only, i.e., $\alpha(x)=0$
for $|x|>1$, $\alpha(x)=\alpha_1$ for $|x|=1$, and arbitrary
$\alpha(0)$, A1 is proved in \cite{HoLa} for $d\geq3$
and A3, A4 are proved in \cite{LuSp} for $d\geq 4$.
The choice of the constants is 
$\gamma=\frac{4}{7}$, $c_2=0$, and the
function $\Fbcr$ is taken to be
\begin{equation}\label{eq:defnnFcr}
 \Fbcr(u;\zeta) = C
 \prod_{\nu=1}^d \frac{1}{|\sin (2 \pi u^\nu)|^{\frac{1}{7}}}
\end{equation}
with a certain constant $C$ depending only on $d$ and $\omega$.
\medskip\hspace*{0cm}\hfill{$\diamondsuit$}\\
\textit{6.4 Main result.} To formulate our main result it is convenient
to integrate the field against a test function $f\in\ell_2$.
Interpreting $\langle\cdot,\cdot\rangle$ as inner product, we set
\begin{equation}\label{5.21}
\langle f,a\rangle= \sum_{x\in\mathbb{Z}^d} f(x)^\ast a(x)\,,\quad
\langle f,a\rangle^\ast=\sum_{x\in\mathbb{Z}^d} f(x)
a(x)^\ast
\end{equation}
and correspondingly for our other conventions as $\langle
f,a(t)\rangle$ and $\langle f,a(\sigma,t)\rangle$. In particular
\begin{equation}\label{5.22}
\langle \hat{f},\hat{a}\rangle=\int_{\mathbb{T}^d}dk
\hat{f}(k)^\ast \hat{a}(k)= \langle f,a\rangle \,.
\end{equation}

Note that by Schwarz inequality for operators and stationarity
\begin{eqnarray}\label{5.23}
&&\hspace{-50pt}\big|\big\langle\langle f_1,a\rangle^\ast\langle
f_2,a(t)\rangle\big\rangle_{\beta,\lambda}\big|^2\leq \big\langle\langle
f_1,a\rangle^\ast\langle f_1,a\rangle\big\rangle_{\beta,\lambda}
\big\langle\langle f_2,a\rangle^\ast\langle
f_2,a\rangle\big\rangle_{\beta,\lambda}
\nonumber\\
&&\hspace{60pt} = \int_{\mathbb{T}^d}dk
W_{\beta,\lambda}^\theta(k)|\hat{f}_1(k)|^2 \int_{\mathbb{T}^d}dk
W_{\beta,\lambda}^\theta(k)|\hat{f}_2(k)|^2\,.
\end{eqnarray}
Hence the quadratic form $(f_1,f_2)\mapsto \big\langle\langle
f_1,a\rangle^\ast\langle f_2,a(t)\rangle\big\rangle_{\beta,\lambda}$ is
uniformly bounded on $\ell_2\times\ell_2$. Our main result concerns
the behavior of this quadratic form on the kinetic time scale.
\medskip\\
\textbf{Theorem 6.4} \textit{Let
$\langle\cdot\rangle_{\beta,\lambda}$ satisfy Assumption A2, let
$\omega$ satisfy Assumptions  A1, A3, and A4, and let $d\geq 4$.
Then there exists $t_0>0$ such that for $0\leq t< t_0$ one has
\begin{equation}\label{5.24}
\lim_{\lambda\to 0}\big\langle
\langle\hat{f}_2,\hat{a}\rangle^\ast
\langle\exp[-\mathrm{i}t\lambda^{-2}\omega^\lambda]
\hat{f}_1,\hat{a}(\lambda^{-2}t)\rangle\big\rangle_{\beta,\lambda}
 = \int_{\mathbb{T}^d}dk
W_{\beta,\lambda}^\theta(k)\hat{f}_1(k)^\ast
\mathrm{e}^{-\nu(k)t}\hat{f}_2(k)
\end{equation}
for all $f_1,f_2\in\ell_2(\mathbb{Z}^d)$.}\medskip\\
\textit{Remark 6.5 (Short kinetic time)}. The restriction to a
short  kinetic time, $t< t_0$, has no physical significance. The
decay is expected to hold for arbitrary $t_0$ and most likely for
even longer times. The smallness of $t_0$ merely reflects that in
our context it is difficult to properly
bound the error terms. Technically it arises because in the error term one 
obtains high order diagrams which are close to the spatially homogeneous 
case as studied in Section \ref{sec.6}. One needs that the leading part 
of the main term is small. Since the leading diagrams are bounded as 
$(t/t_0)^N$, one has to require $t < t_0$. \medskip\hspace*{0cm}\hfill{$\diamondsuit$}\\
\textit{6.5 Link to the nonlinear Schr\"{o}dinger equation as discussed
in \cite{LuSp}}. \cite{LuSp} is written in such a way that the
operator ordering, as of relevance in our context, is already
respected. Thus, with the proper reinterpretation, most formulas
hold also for the quantum evolution. In particular, the analysis of
the Feynman diagrams carries over \textit{verbatim}. The Duhamel
expansion (\ref{2.20}) is preliminary and one still needs three
modifications to reach the starting point of \cite{LuSp}.\medskip\\
$-$ partial time integration. This refers to the time integrations and
thus remains valid in the quantum mechanical context.\medskip\\
$-$ insertion of the cutoff functions
  $\Phi^\lambda_0,\Phi^\lambda_1$. This refers to the $k$-integrations and
  thus remains valid in the quantum mechanical context.\medskip\\
$-$ removal of fast oscillations. As argued above the
  dispersion relation $\omega$ is renormalized to $\omega^\lambda$.
  Thus, as a modification of (\ref{2.9d}), we define
  \begin{equation}\label{5.24a}
  \hat{a}(k,1,t)= e^{\mathrm{i}\omega^\lambda(k)t}
  \hat{a}(k,t)\,,\quad
  \hat{a}(k,-1,t)= e^{-\mathrm{i}\omega^\lambda(k)t}
  \hat{a}(-k,t)^\ast\,.
  \end{equation}
In addition we introduce the pair truncation
\begin{eqnarray}\label{5.24b}
&&\hspace{-40pt}\hat{\mathcal{P}}\big(\hat{a}(k_1,-1)\hat{a}(k_2,\sigma)\hat{a}(k_3, 1)\big)
\nonumber\\
&&\hspace{-20pt}=\hat{a}(k_1,-1)\hat{a}(k_2,\sigma)\hat{a}(k_3, 1)-\big(\langle
\hat{a}(k_1,-1)\hat{a}(k_2,\sigma)\rangle_{\beta,\lambda}\hat{a}(k_3,1)\nonumber\\
&&\hspace{-10pt}+ \theta\langle \hat{a}(k_1,-1)
\hat{a}(k_3,1)\rangle_{\beta,\lambda}\hat{a}(k_2,\sigma) + 
\langle \hat{a}(k_2,\sigma)
\hat{a}(k_3,1)\rangle_{\beta,\lambda}\hat{a}(k_1,-1)\big)\,,
\end{eqnarray}
$\sigma = \pm 1$ and omitting the $t$ argument. Then  $\hat{a}(k_1,\sigma,t)$
satisfies the evolution equation 
\begin{eqnarray}\label{5.24c}
&&\hspace{-50pt}\frac{d}{dt}\hat{a}(k_1,\sigma,t)=-\mathrm{i}\lambda\sigma
\int_{(\mathbb{T}^d)^3} dk_2 dk_3 dk_4 \delta(k_1-k_2-k_3-k_4)\nonumber\\
&&\hspace{25pt}\times\tfrac{1}{2}\big((1+\sigma)\widehat{V}(k_2+k_3)
+(1-\sigma)\widehat{V}(k_3+k_4)\big)\nonumber\\
&&\hspace{25pt}\times \textrm{e}^{-\mathrm{i}t\Omega(k,\sigma)}
\big\{\Phi^\lambda_1(k_2,k_3,k_4)\hat{a}(k_2,-1,t)
\hat{a}(k_3,\sigma,t)\hat{a}(k_4,1,t)\nonumber\\
&&\hspace{32pt} +
\Phi^\lambda_0(k_2,k_3,k_4)\mathcal{\hat{P}}[\hat{a}(k_2,-1,t)
\hat{a}(k_3,\sigma,t) \hat{a}(k_4,1,t)]\big\}\,,
\end{eqnarray}
where
\begin{equation}\label{5.24d}
    \Omega(k,\sigma)=-\sigma
    \omega^\lambda(k_1)-\omega^\lambda(k_2)+\sigma\omega^\lambda(k_3)
    +\omega^\lambda(k_4)\,.
\end{equation}
(\ref{5.24c}) agrees with the corresponding formula of \cite{LuSp}
with the only modification consisting of a general interaction potential. In
\cite{LuSp} we consider the case $\hat{V}(k)=\hat{V}(0)$. Therefore
the factor  in (\ref{5.24c}) containing $\widehat{V}$ is replaced by a constant and
$\omega^\lambda(k)=\omega(k)+\lambda R_0$, where $R_0$ depends on
$\lambda$ but not on $k$. In frequency differences $R_0$ thus drops
\medskip out.

We are confident that the estimates of the oscillatory integrals
proved in \cite{LuSp} still hold for the case under consideration
here. Of course, a simple estimate as
\begin{equation}\label{5.24e}
|\mathrm{e}^{-\textrm{i}\omega(k)t}-\mathrm{e}^{-\textrm{i}\omega^\lambda(k)t}|\leq
C\lambda|t|
\end{equation}
uniformly in $k$ is too crude. Rather each oscillating integral has
to be estimated with an $\omega^\lambda$ depending weakly on
$\lambda$. This step remains to be carried out.

The classical average $\mathbb{E}$ in \cite{LuSp}  becomes the quantum average
$\langle\cdot\rangle_{\beta,\lambda}$ over the KMS state. With the
assumption A2, higher cumulants do not contribute. For the
quasifree part one has either $W$ or $\tilde{W}$ as time $t=0$
input. Since both functions are smooth, the estimates from the
commutative case remain valid.\medskip\\
\textit{6.6 Error terms.} We adopt the notation from \cite{LuSp}. There
are three error terms of the same structure, namely
\begin{equation}\label{5.35}
\big\langle\langle\hat{f}_1,\hat{a}(0)\rangle^\ast\int^t_0 ds
\langle
\hat{f}_2,\mathcal{X}_n(t,s)[\hat{a}(s)]\rangle\big\rangle_{\beta,\lambda}
\end{equation}
with $\mathcal{X}$ any of $\mathcal{G}$, $\mathcal{Z}$, or
$\mathcal{A}$. We use the Schwarz inequality for operators $A,B$
according to which
\begin{equation}\label{5.36}
|\langle A^\ast B\rangle_{\beta,\lambda}|^2\leq \langle |A|^2
\rangle_{\beta,\lambda} \langle ]B|^2\rangle_{\beta,\lambda}
\end{equation}
with the shorthand $A^\ast A = |A|^2$. Then
\begin{eqnarray}\label{5.36a}
&&\hspace{-40pt}
\Big|\big\langle\langle\hat{f}_1,\hat{a}(0)\rangle^\ast
\langle\hat{f}_2,\int^t_0 ds
\mathcal{X}_n(s,t)[\hat{a}(s)]\rangle\big\rangle_{\beta,\lambda}\Big|^2\nonumber\\
&&\hspace{-20pt}\leq\Big(\int^t_0 ds\big|\big\langle
\langle\hat{f}_1,\hat{a}(0)\rangle 
\langle\hat{f}_2,
\mathcal{X}_n(s,t)[\hat{a}(s)]\rangle\big\rangle_{\beta,\lambda}\big|\Big)^2\nonumber\\
&&\hspace{-20pt}\leq t
\big\langle|\langle\hat{f}_1,\hat{a}(0)\rangle|^2\big\rangle_{\beta,\lambda}
\int^t_0 ds \big\langle|\langle\hat{f}_2,
\mathcal{X}_n(t,s)[\hat{a}(s)]\rangle|^2\big\rangle_{\beta,\lambda}\,.
\end{eqnarray}
$\mathcal{X}_n(t,s)$ is a monomial
of order $n$ in the factors $a(k,\sigma,s)$ which differ by a phase
factor from $a(k,s)$, $a(k,s)^\ast$. The latter field operators are
invariant for $\langle\cdot\rangle_{\beta,\lambda}$ and,  at the
expense of a phase factor, $\hat{a}(s)$ in (\ref{5.36a}) can be
replaced by $\hat{a}(0)$. Thus, at given cutoff $N$, the two-point
function of (\ref{5.24}) is reduced to a main term and an error
term, which both involve only the free time evolution and are
monomials of the time 0 fields $\hat{a}$ averaged
with respect to the $\beta$-KMS state $\langle\cdot\rangle_{\beta,\lambda}$.\medskip\\
\textit{6.7 Convergence of the leading part of the main term.} We refer
to Section \ref{sec.6} for the general structure. For the case under
consideration the Feynman diagrams are constrained. In (\ref{5.24})
the oscillating term is combined with
$\langle\hat{f}_2,\hat{a}\rangle^\ast$. The Feynman diagram has 
one line segment $[0,t]$ with parity $-1$, which is constrained not 
to fuse at all. In the time slice $[t-s_n,t]$ there is only one further 
line segment. Thus $n_0 = 2$. A leading diagram is still defined
by the pairing property for even time slices and the absence of the
factor $\widehat{V}(0)$. The line corresponding to
$\langle\hat{f}_2,\hat{a}\rangle^\ast$ has momentum $k_1$. The
collision operator acting on $k_2,\ldots,k_{n+2}$ is defined as in
(\ref{6.3}). Because of the constraint the collision operator acting
on $k_1$ reads
\begin{eqnarray}\label{5.38}
&&\hspace{-25pt}\big(\mathcal{C}^-_{1,n+2}(t)\widehat{\rho}_{n+2}\big)
(k_1,\tau_1,\ldots,k_n,\tau_n)=2\int_{(\mathbb{T}^d)^3}
dg_2 dg_3 dg_4 \delta(k_1+g_2-g_3-g_4)\nonumber\\
&&\hspace{10pt}\times\mathrm{e}^{\mathrm{i}t(\omega(k_1)+\omega(g_2)-\omega(g_3)-\omega(g_4))}
\widehat{V}(k_2-k_3)\big(\widehat{V}(k_2-k_3)+\theta\widehat{V}(k_2-k_4)\big)\nonumber\\
&&\hspace{10pt}\times\big(\theta\widehat{\rho}_{n+2}(k_1,\tau_1,\ldots,g_3,1,g_4,1)-
\widehat{\rho}_{n+1}(k_1,\tau_1,\ldots,g_2,1,g_3,-1)
\nonumber\\
&&\hspace{20pt}-\theta\widehat{\rho}_{n+2}(k_1,\tau_1,\ldots,g_2,1,g_4,1)\big)\,.
\end{eqnarray}
The initial conditions are
\begin{equation}\label{5.39}
\widehat{\rho}_n(k_1,\tau_1,\ldots,k_n,\tau_n)=
\prod^n_{j=1}W_\beta(k_j,\tau_j)\,.
\end{equation}
We set
\begin{equation}\label{5.40}
    \mathcal{C}^-_{n+2}(t)=\mathcal{C}^-_{1,n+2}(t)+ \sum^n_{j=2}
    \mathcal{C}_{j,n+2}(t)\,.
\end{equation}

Let us define $\mathcal{M}^\lambda_n(t)$ as the sum of all leading
Feynman diagrams at order $n$. With the above notation it is given
by
\begin{eqnarray}\label{5.41}
&&\hspace{-30pt}\mathcal{M}^\lambda_n(t)=
\lambda^{2n}\int_{(\mathbb{R}_+)^{n+1}}d\underline{t}
\int_{(\mathbb{R}_+)^n}d\underline{s}\delta(\sum^{n+1}_{j=1}
t_j+\sum^n_{j=1} s_j-t)\int_{\mathbb{T}^d}
dk_1\hat{f}_1(k_1)^\ast \hat{f}_2(k_1)\nonumber\\
&&\hspace{24pt}\big( \mathcal{C}^-_3(s_1)\ldots
\mathcal{C}^-_{2n+1}(s_n)\widehat{\rho}_{2n+1}\big)(k_1,1)\,.
\end{eqnarray}
Proposition 5.7 remains valid. Thus one can pass to the limit as
\begin{equation}\label{5.42}
\lim_{\lambda\to 0}\mathcal{M}^\lambda_n(\lambda^{-2}t)=
\frac{1}{n!} t^n\int_{\mathbb{T}^d} dk_1
\hat{f}_1(k_1)^\ast\hat{f}_2(k_1)
(\mathcal{C}^-_3\ldots\mathcal{C}^-_{2n+1}\widehat{\rho}_{2n+1})(k_1,1)\,.
\end{equation}

Let $\widehat{\rho}_3$ be given by (\ref{5.39}). Then
\begin{eqnarray}\label{5.43}
&&\hspace{-40pt}\big(\mathcal{C}_{1,3}(s)\widehat{\rho}_{3}\big)
(k_1,\tau_1)= \frac{d}{ds}\int_{(\mathbb{T}^d)^3}
dk_2 dk_3 dk_4 \delta(k_1+k_2-k_3-k_4)\nonumber\\
&&\hspace{40pt}\times(\sin\Omega s)\Omega^{-1}(\mathrm{e}^{\beta\Omega}-1)
\big(\widehat{V}(k_2-k_3)+\theta(\widehat{V}(k_2-k_4)\big)^2\nonumber\\
&&\hspace{40pt}\times W_\beta(k_1,\tau_1)W_\beta(k_2,\tau_1)
W_\beta(k_3,-\tau_1)W_\beta(k_4,-\tau_1)\,,
\end{eqnarray}
with $\Omega=\omega_1+\omega_2-\omega_3-\omega_4$. Therefore
\begin{equation}\label{5.44}
\int^\infty_0 ds \big(\mathcal{C}_{1,3}(s)
\widehat{\rho}_3\big)(k_1,\tau_1)=0\,.
\end{equation}
Returning to (\ref{5.42}) we conclude that when acting with
$\mathcal{C}^-_{2j+1}$, where $j=1,\ldots,n$, only its first
summand, i.e., $\mathcal{C}^-_{1,2j+1}$, contributes. Hence, there exists $t_0$
such that for $0 \leq t \leq t_0$ one has 
\begin{equation}\label{5.45}
\lim_{\lambda\to
0}\sum^\infty_{n=0}\mathcal{M}^\lambda_n(\lambda^{-2}t)=
\sum^\infty_{n=0}\frac{t^n}{n!} (-1)^n\int_{\mathbb{T}^d} dk
\hat{f}_1(k)^\ast\hat{f}_2(k)W_{\beta}^\theta(k)\nu(k)^n\,.
\smallskip
\end{equation}
\textit{Remark 6.6 (Radius of convergence)}. The integral in
(\ref{5.43}) decays at least as $s^{-\delta}$. Therefore the error
in (\ref{5.44}) is order $\lambda^{2\delta}$. The leading diagrams
at order $n$ have a single term which yields the non-zero limit in
(\ref{5.45}), while the remaining $(n!-1)$ terms are of order
$\lambda^{2\delta}$. Therefore one can cut the series at order $N$
with $\lambda^{2\delta} N!=1$. Since the series of contributing
terms has an infinite radius of convergence, the restriction to
bounded $t_0$ can be lifted and (\ref{5.45}) holds in fact for all
$t$. On the other hand the error term can be controlled only for
$|t|\leq t_0$ and the extra effort will not improve Theorem
6.3.\hspace*{0cm}\hfill{$\diamondsuit$}

%%%%%%%%%%%%%%%%%%%%%%%%%%%%%%%%%%%%%%%%%%%%%%%%%%%%%%%%%%%%%%%%

\section{Time correlations for the number density}\label{sec.7}
\setcounter{equation}{0}

Physically of great interest are the density current and energy
current time correlations. They differ in two respects from the
correlation function studied in the previous section: they are
four-point functions and, more importantly, involve a spatial
summation. Slightly formal, these correlation functions are
particular cases of the number density time correlations in
equilibrium, i.e., of $\langle \hat{a}(k,t)^\ast \hat{a}(k,t) 
\hat{a}(k',0)^\ast
\hat{a}(k',0)\rangle_{\beta,\lambda}$.

Let $\eta:\mathbb{Z}^d\to\mathbb{C}$ such that
$\eta(x)^\ast=\eta(-x)$ and $\eta$ has bounded support. We define
the energy-like, resp.\ number-like, observable
\begin{equation}\label{7.1}
H^\eta=\sum_{x,y\in\mathbb{Z}^d}\eta(x-y)a(x)^\ast a(y)
\end{equation}
and represent it by a sum of local terms as
\begin{eqnarray}\label{7.2}
&&\hspace{-10pt}H^\eta=\sum_{w\in\mathbb{Z}^d}H_w^\eta\,,\nonumber\\
&&\hspace{-10pt}H^\eta_w=\tfrac{1}{2}\sum_{y\in\mathbb{Z}^d}\big(a(w)^\ast\eta(w-y)a(y)
+a(y)^\ast\eta(y-w)a(w)\big)\,.
\end{eqnarray}
For $H^\eta=H_\mathrm{har}$ one has $\hat{\eta}(k)=\omega(k)$,
for the energy current
$\hat{\eta}(k)=(\nabla_k\omega(k))\omega(k)$, and,
correspondingly, $\hat{\eta}(k)=1$,
$\hat{\eta}(k)=\nabla_k\omega(k)$ for number and number current.

As before $\langle\cdot\rangle_{\beta,\lambda}$ is the $\beta$-KMS
state for $H-\mu N$ at infinite volume with $\mu$ satisfying
(\ref{5.1}) and we set
$\langle\cdot\rangle_{\beta,0}=\langle\cdot\rangle_{\beta}$. It is
also convenient to introduce the Kubo inner product defined
through
\begin{equation}\label{7.2a}
\langle \!\langle A,B \rangle\!\rangle_{\beta,\lambda}=
\beta^{-1}\int^\beta_0 d\beta'\big(\langle A^\ast e^{-\beta'H} B
e^{\beta'H}\rangle_{\beta,\lambda} - \langle
A^\ast\rangle_{\beta,\lambda}\langle
B\rangle_{\beta,\lambda}\big)\,.
\end{equation}
The fluctuations of $H^\eta$ in a large box $\Lambda$ are given by
\begin{equation}\label{7.3}
\xi^\eta_\Lambda=|\Lambda|^{-1/2}\sum_{w\in
\Lambda}\big(H^\eta_w-\langle H^\eta_w\rangle_{\beta,\lambda}\big)
\end{equation}
and the quantity of interest is the time-displaced covariance of
(\ref{7.3}), which reads
\begin{equation}\label{7.5}
\lim_{\Lambda\uparrow\mathbb{Z}^d}
\langle\!\langle\xi^\eta_\Lambda,\xi^\eta_\Lambda(t)\rangle\!\rangle_{\beta,\lambda}
=\sum_{w\in\mathbb{Z}^d} \langle\!\langle H^\eta_w,
H^\eta_0(t)\rangle\!\rangle_{\beta,\lambda} = C^\eta_\lambda(t)\,.
\end{equation}
In particular, at $t=0$,
\begin{equation}\label{7.4}
\lim_{\lambda\to
0}\lim_{\Lambda\uparrow\mathbb{Z}^d} \langle\!\langle
\xi^\eta_\Lambda,\xi^\eta_\Lambda
\rangle\!\rangle_{\beta,\lambda}=\sum_{w\in\mathbb{Z}^d}
\langle\!\langle H^\eta_w, H^\eta_0\rangle\!\rangle_\beta
=\int_{\mathbb{T}^d} dk |\hat{\eta}(k)|^2
W^\theta_\beta(k) \tilde{W}^\theta_\beta(k)\,.
\end{equation}
Note that in the limit $\lambda\to 0$ the Kubo and standard inner
product coincide for the observables under consideration.

As before the kinetic limit provides information on
$C^\eta_\lambda(\lambda^{-2}t)$ for small $\lambda$. There is a
simple formal argument how to guess this correlation by building on
the results from Section \ref{sec.6}. Let
$\langle\cdot\rangle_{\beta,\lambda}(\varepsilon)$ be
the $\beta$-KMS state for $H-\mu N+(\varepsilon/\beta)H^\eta$. Then
\begin{equation}\label{7.6}
C^{\eta}_{\lambda}(t)=\frac{\partial}{\partial\varepsilon} \langle
H^\eta_0(t)\rangle_{\beta,\lambda}(\varepsilon)\big|_{\varepsilon=0}\,.
\end{equation}
We now perform first the limit $\lambda\to 0$ and then the limit
$\varepsilon\to 0$. For time $t=0$ one has
\begin{equation}\label{7.7}
\lim_{\lambda\to 0}\langle \hat{a}(k)^\ast
\hat{a}(k')\rangle_{\beta,\lambda}(\varepsilon)= \delta(k-k')
W^\varepsilon(k)\,,
\end{equation}
where
\begin{equation}\label{7.7a}
W^\varepsilon(k)=
(\mathrm{e}^{[\beta(\omega(k)-\mu)+\varepsilon\hat{\eta}(k)]}
-\theta)^{-1}\,,
\end{equation}
which is well-defined for $\varepsilon$ sufficiently small. Now, by
Section \ref{sec.6},
\begin{equation}\label{7.8}
\lim_{\lambda\to 0}\langle \hat{a}(k,\lambda^{-2}t)^\ast
\hat{a}(k',\lambda^{-2}t)\rangle_{\beta,\lambda}(\varepsilon)=
\delta(k-k')W^\varepsilon(k,t)\,,
\end{equation}
and $W^\varepsilon(k,t)$ solves (\ref{6a.4}) with initial condition
$W^\varepsilon(k)$. It follows that
\begin{equation}\label{7.9}
\lim_{\lambda\to 0}\langle H^\eta_0 (\lambda^{-2}t)
\rangle_{\beta,\lambda}(\varepsilon)= \int_{\mathbb{T}^d} dk
\hat{\eta}(k) W^\varepsilon(k,t)\,.
\end{equation}

We linearize the collision operator as
\begin{equation}\label{7.10}
\mathcal{C}(W^\theta_\beta + \varepsilon f)=\varepsilon
Af+\mathcal{O}(\varepsilon^2)\,.
\end{equation}
It is convenient to introduce the multiplication operator $U_\beta$
through
\begin{equation}\label{7.10a}
(U_\beta f)(k) = \big(W^\theta_\beta(k)
\tilde{W}^\theta_\beta(k)\big)^{1/2} f(k)\,,
\end{equation}
and to define
\begin{equation}\label{7.10b}
Lf =- A U_\beta^2  f\,.
\end{equation}
Expanding (\ref{7.9}),
\begin{equation}\label{7.11}
\frac{\partial}{\partial\varepsilon}\int_{\mathbb{T}^d} dk
\hat{\eta}(k) W^\varepsilon(k,t)\big|_{\varepsilon=0}= \langle
\hat{\eta},\mathrm{e}^{At}U_\beta^2 \hat{\eta}\rangle\,,
\end{equation}
where $\langle\cdot,\cdot\rangle$ denotes now the inner product in
$L^2(\mathbb{T}^d, dk)$. Using the definition of $L$, (\ref{7.11})
can be written in the more symmetric form as
\begin{equation}\label{7.12}
 \langle\hat{\eta},\mathrm{e}^{At}U_\beta^2
\widehat{ \eta}\rangle= \langle U_\beta
 \hat{\eta}, \exp[-U_\beta^{-1}
 L U_\beta^{-1} t] U_\beta \hat{\eta}\rangle\,.
\end{equation}

From the linearization one obtains
\begin{eqnarray}\label{7.13}
&&\hspace{-46pt}(Lf)(k_1)= \pi \int_{(\mathbb{T}^d)^3} dk_2 dk_3
dk_4 \delta(k_1+k_2-k_3-k_4)\delta(\omega_1
+\omega_2-\omega_3-\omega_4)\nonumber\\
&&\hspace{-10pt}\times|\widehat{V}(k_2-k_3)+\theta
\widehat{V}(k_2-k_4)|^2 W^\theta_{\beta,1}W^\theta_{\beta,2}
\tilde{W}^\theta_{\beta,3}\tilde{W}^\theta_{\beta,4}
(f_1+f_2-f_3-f_4)\,.
\end{eqnarray}
The quadratic form associated to $L$  reads then
\begin{eqnarray}\label{7.14}
&&\hspace{-40pt}\langle f,Lf\rangle= \frac{\pi}{4}
\int_{(\mathbb{T}^d)^4} dk_1 dk_2 dk_3 dk_4
\delta(k_1+k_2-k_3-k_4)\delta(\omega_1
+\omega_2-\omega_3-\omega_4)\nonumber\\
&&\hspace{-10pt}\times|\widehat{V}(k_2-k_3)+\theta
\widehat{V}(k_2-k_4)|^2 W^\theta_{\beta,1}W^\theta_{\beta,2}
\tilde{W}^\theta_{\beta,3}\tilde{W}^\theta_{\beta,4}
(f_1+f_2-f_3-f_4)^2\,.
\end{eqnarray}
Therefore $L=L^\ast$ and $L\geq 0$. Clearly $L1=0$ and $L\omega=0$.
Thus the zero subspace is at least two-fold degenerate. If the
H-theorem holds, see Appendix B, then the 0 eigenvalue of $L$ is
exactly two-fold degenerate. In brackets, we note that momentum
conservation is destroyed by the underlying lattice. More precisely,
if one sets $f(k)=k$ mod 1, then $Lf\neq0$, since the umklapp
$k_1+k_2-k_3-k_4=n$, $n$ integer vector and $n\neq 0$, is permitted
according to the momentum $\delta$-function.

Allowing for the interchange of limits we arrive at\medskip\\
\textbf{Conjecture 7.1} \textit{Under suitable conditions on $\omega$ it
holds}
\begin{equation}\label{7.15} \lim_{\lambda\to
0} C^\eta_\lambda(\lambda^{-2}t)= \langle U_\beta \hat{\eta},
\exp[- U_\beta^{-1} LU_\beta^{-1} |t|] U_\beta
\hat{\eta}\rangle\,.\medskip
\end{equation}
\textit{Duhamel expansion.} We want to explore whether, at least in
principle, the Duhamel expansion of Section \ref{sec.2} could work.
The starting point is (\ref{2.20}) for $n_0=2$. This results in
\begin{equation}\label{7.16}
H^\eta_0(t)=H^\eta_{0,\mathrm{main}}(t)+H^\eta_{0,\mathrm{error}}(t)\,.
\end{equation}
The main term is discussed first.

We insert the main term in the definition of $C^\eta_\lambda(t)$.
Then the time $t=0$ input for the Feynman diagrams is of the generic
form
\begin{equation}\label{7.17}
\sum_{w\in\mathbb{Z}^d} \langle\!\langle H^\eta_w, \prod^n_{j=1}
\hat{a}(k_j,\sigma_j)\rangle\!\rangle_{\beta,\lambda}
=\frac{\partial}{\partial\varepsilon}\langle\prod^n_{j=1}
\hat{a}(k_j,\sigma_j)\rangle_{\beta,\lambda}(\varepsilon)\big|_{\varepsilon=0}\,.
\end{equation}
Therefore we have to require a slightly modified $\ell_1$-clustering
as\medskip\\
 \textbf{Assumption A5.} (strengthened $\ell_1$-clustering). \textit{Let
$\beta>0$ and $\mu$ satisfy (\ref{5.1}). There exists $\lambda_0>0$
and $c_0>0$ independent of $n$ such that for $0<\lambda\leq
\lambda_0$ and all $n\geq 4$ one has the bound
\begin{equation}\label{7.18}
\sum_{x\in(\mathbb{Z}^d)^n}
\delta_{x_1 0}\Big|\frac{\partial}{\partial\varepsilon}\langle
\prod^n_{j=1}
a(x_j,\sigma_j)\rangle_{\beta,\lambda}(\varepsilon)\big|_{\varepsilon=0}\Big|\leq
\lambda(c_0)^n n!\,.
\end{equation}
In addition the following limit exists uniformly in $k$}
\begin{equation}\label{7.18a}
\lim_{\lambda\to 0}\sum_{x\in \mathbb{Z}}\Big|
\frac{\partial}{\partial\varepsilon}\langle a(x)^\ast
a(0)\rangle_{\beta,\lambda}(\varepsilon)\big|_{\varepsilon=0}
- (U^2_\beta\hat{\eta})\,\check{}\,(x)\Big|= 0\,.\medskip
\end{equation}

Under Assumption A5, the input function for the Feynman diagrams
has the same regularity as used in Section \ref{sec.5}. We conclude
that together with Assumptions A1, A3, and A4, the Duhamel
expansion converges term to the conjectured limit (\ref{7.15}).

For the error term one has symbolically
\begin{equation}\label{7.19}
H^\eta_{w,\mathrm{error}}(t)=\int^t_0 ds
\mathcal{F}^\eta_{N,w}(t-s)[\hat{a}(s)]\,.
\end{equation}
For the Kubo inner product we use Schwarz inequality as in
(\ref{5.36}) to conclude
\begin{eqnarray}\label{7.20}
&&\hspace{-40pt}\Big(\sum_{w\in\mathbb{Z}^d} \langle\!\langle
H^\eta_w, H^\eta_{0,\mathrm{error}}(t)
\rangle\!\rangle_{\beta,\lambda} \Big)^2\nonumber\\
&&\hspace{-32pt}\leq t \big(\sum_{w\in\mathbb{Z}^d} \langle\!\langle
H^\eta_0, H^\eta_w \rangle\!\rangle_{\beta,\lambda}\big) \int^t_0
ds \sum_{w\in\mathbb{Z}^d}\langle\!\langle
\mathcal{F}^\eta_{N,0}(t-s)[\hat{a}(s)],
\mathcal{F}^\eta_{N,w}(t-s)[\hat{a}(s)]\rangle\!\rangle_{\beta,\lambda}
\,.
\end{eqnarray}
The first factor is bounded. In the second factor we use stationary
to arrive at
\begin{equation}\label{7.21}
\int^t_0 ds \sum_{w\in\mathbb{Z}^d} \langle\!\langle
\mathcal{F}_{N,0}(s)[\hat{a}],
\mathcal{F}_{N,w}(s)[\hat{a}]\rangle\!\rangle_{\beta,\lambda} \,.
\end{equation}
Proceeding as before, one would like to take the $\sum_w$ in the
exponential to modify the state
$\langle\cdot\rangle_{\beta,\lambda}$. But this would be a high
order polynomial, hence difficult to control. One may view
(\ref{7.21}) also as Feynman diagrams with singular initial
conditions resulting from the one $\delta$-function because of the
$\sum_w$. This would require to redo the oscillatory integrals. At
present it is not clear whether such an approach could work even in
principle.

%%%%%%%%%%%%%%%%%%%%%%%%%%%%%%%%%%%%%%%%%%%%%%%%%%%%%%%%%%%%%%%%%%%%

\begin{appendix}
\section{Appendix. Truncated correlation functions}\label{sec.A}
\setcounter{equation}{0}

We follow Bratelli and Robinson \cite{BrRo}, pages 39 and 43.

Let $\langle\cdot\rangle$ be an even state on the CCR, resp.\ CAR,
algebra. The moments are assumed to exist and odd moments vanish. We
use $a_j$ as shorthand for $\langle f_j,a_j(\sigma_j)\rangle$. Then
the moments are
\begin{equation}\label{A.1}
\langle \prod^n_{j=1}a_j\rangle
\end{equation}
and the fully truncated moments are denoted by
\begin{equation}\label{A.2}
\langle \prod^n_{j=1}a_j\rangle^T\,.
\end{equation}

To define them, let $\mathfrak{J}$ denote an index set and $F$ a
function from the non empty ordered subsets of $\mathfrak{J}$ to the
complex numbers. The truncation $F_T$ of $F$ is now given
recursively by
\begin{equation}\label{A.3}
F(I)=\sum_{\mathcal{P}_I}\varepsilon
(\mathcal{P}_I)\prod_{J\in\mathcal{P}_I} F_T(J)\,,
\end{equation}
where the sum is over all partitions $\mathcal{P}_I$ of $I$ into
ordered even subsets, $\mathcal{P}_I=\{J_1,\ldots,J_n\}$ and
$\varepsilon(\mathcal{P}_I)=1$ for bosons and
$\varepsilon(\mathcal{P}_I)=\pm 1$ is according to whether the
permutation $I\mapsto(J_1,\ldots,J_n)$ is even or odd. In our
case $F$ are the moments and $F_T$ their full truncation.\medskip\\
\textbf{Theorem A.1} (truncated correlations). \textit{Let $n$ be even, $n\geq 4$, and let
$\pi$ be a permutation of $I=(1,\ldots,n)$. Then
\begin{equation}\label{A.4}
\langle \prod^n_{j=1} a_j\rangle^T=\varepsilon(\pi)\langle
\prod^n_{j=1} a_{\pi(j)}\rangle^T\,,
\end{equation}
where $\varepsilon(\pi)=1$ for bosons and for fermions
$\varepsilon(\pi)=\mathrm{sign}\pi$, the sign of the permutation $\pi$.}\medskip\\
\begin{Proof} Let $I=\{1,\ldots,n\}$ and
$\alpha=\{m,m+1\}\subset I$. We set
\begin{equation}\label{A.5}
F(I)=\langle a_1\ldots a_n\rangle\,,\quad \widetilde{F}(I)=\langle
a_1\ldots a_{m+1} a_m \ldots a_n\rangle\,.
\end{equation}
It holds
\begin{equation}\label{A.5a}
F(I)=\sum_{\alpha\in J\subset I}\varepsilon(J,I\smallsetminus
J)F_T(J)F(I\smallsetminus J)\,,
\end{equation}
where $\varepsilon(J,I\smallsetminus J)=1$ for bosons and
$\varepsilon(J,I\smallsetminus J)=\pm 1$ for fermions according to
whether $I\mapsto (J,I\smallsetminus J)$ is an even or odd
permutation of $I$. Since $a_m a_{m+1}-\theta a_{m+1}a_m= c
\mathbbm{1}$, $c\in\mathbb{C}$, one has
\begin{eqnarray}\label{A.6}
&&\hspace{-10pt}c F(I\smallsetminus \alpha)=
F(I)-\theta\widetilde{F}(I)=\sum_{\alpha\in J\subset
I}\varepsilon(J,I\smallsetminus J)\big(F_T(J)-
\theta\widetilde{F}_T(J)\big)F(I\smallsetminus J)\\
&&\hspace{-10pt}=\big(F_T(\alpha)-
\theta\widetilde{F}_T(\alpha)\big)F(I\smallsetminus\alpha)
+\sum_{\alpha\in J\subset I,J \neq \alpha}
\varepsilon(J,I\smallsetminus J)\big(F_T(J)-
\theta\widetilde{F}_T(J)\big)F(I\smallsetminus J)\nonumber\,.
\end{eqnarray}
Since $F_T(\alpha)- \theta\widetilde{F}_T(\alpha)=c$, we conclude
\begin{equation}\label{A.7}
0=\sum_{\alpha\in J\subset I, J \neq \alpha}
\varepsilon(J,I\smallsetminus J)\big(F_T(J)-
\theta\widetilde{F}_T(J)\big)F(I\smallsetminus J)\,.
\end{equation}
For $|I|=4$, the only summand is $J=I$ and
$F_T(I)=\theta\widetilde{F}_T(I)$ since $F(\emptyset)=1$. By
iterating (\ref{A.7}), one concludes the validity of (\ref{A.4}).
\end{Proof}

\section{Appendix. Some properties of the spatially homogeneous
Boltzmann-Nordheim equation}\label{sec.B} \setcounter{equation}{0}

The purpose of this appendix is to provide a rather compressed list
of the basic properties of the spatially homogeneous
Boltzmann-Nordheim equation and to point at the relevant literature.
It will be convenient to discuss fermions and bosons separately.

\subsection{Fermions}\label{B.1}

The Boltzmann-Nordheim equation reads, see (\ref{6a.4}),
(\ref{6a.5}),
\begin{equation}\label{B.1a}
\frac{\partial}{\partial t} W(k,t)=\mathcal{C}\big(W(t)\big)(k)\,,
\end{equation}
\begin{eqnarray}\label{B.2a}
&&\hspace{-53pt}\mathcal{C}(W)(k_1)= \pi \int_{(\mathbb{T}^d)^3}
dk_2 dk_3 dk_4 \delta(k_1+k_2-k_3-k_4)\delta(\omega_1
+\omega_2-\omega_3-\omega_4)\nonumber\\
&&\hspace{15pt}|\widehat{V}(k_2-k_3)-\widehat{V}(k_2-k_4)|^2
(\tilde{W}_1\tilde{W}_2 W_3 W_4-W_1 W_2\tilde{W}_3\tilde{W}_4)\,,
\end{eqnarray}
$\tilde{W}=1-W$. It has to be solved for initial data $W$ such that
$0\leq W(k)\leq 1$. If there would be a first time $t$ such that
$W(k_0,t)=0$, then $\mathcal{C}\big(W(t)\big)(k_0)\geq 0$, and
correspondingly $W(k_1,t)=1$ implies that
$\mathcal{C}\big(W(t)\big)(k_1)\leq 0$. Hence the constraint $0\leq
W\leq 1$ is preserved in time. Note that $\mathcal{C}(W)=0$ in case
$\widehat{V}(k)=\widehat{V}(0)$.

At least formally, total number and energy are conserved,
\begin{equation}\label{B.3}
\int_{\mathbb{T}^d} dk_1 \mathcal{C}(W)(k_1)=0\,,\quad
\int_{\mathbb{T}^d} dk_1 \omega(k_1) \mathcal{C}(W)(k_1)=0\,.
\end{equation}
According to the Fermi statistics, the entropy per volume of a
quasifree state is given by
\begin{equation}\label{B.4}
S(W)=-\int_{\mathbb{T}^d} dk \big(W(k)\log W(k)+
\tilde{W}(k)\log\tilde{W}(k)\big)\,.
\end{equation}
The entropy changes in time as
\begin{equation}\label{B.5}
\frac{d}{dt}S\big(W(t)\big)=\sigma\big(W(t)\big)
\end{equation}
with the entropy production
\begin{eqnarray}\label{B.6}
&&\hspace{-40pt}\sigma(W)= \pi \int_{(\mathbb{T}^d)^4} dk_1 dk_2
dk_3 dk_4 \delta(k_1+k_2-k_3-k_4)\delta(\omega_1
+\omega_2-\omega_3-\omega_4)\nonumber\\
&&\hspace{8pt}|\widehat{V}(k_2-k_3)- \widehat{V}(k_2-k_4)|^2
F(\tilde{W}_1\tilde{W}_2 W_3 W_4,W_1 W_2\tilde{W}_3\tilde{W}_4)
\end{eqnarray}
and $F(x,y)=(x-y)\log (x/y)$. Hence
\begin{equation}\label{B.7}
\sigma\geq 0\,,
\end{equation}
which is the H-theorem. $\sigma=0$ if and only if
\begin{equation}\label{B.8}
\frac{W_1W_2}{\tilde{W}_1\tilde{W}_2}=
\frac{W_3W_4}{\tilde{W}_3\tilde{W}_4}
\end{equation}
on the collision set $\mathcal{D}_{\mathrm{c}}=\{(k_1,k_2,k_3,k_4)\in
\mathbb{R}^{12}| k_1+k_2=k_3+k_4,
\omega_1+\omega_2=\omega_3+\omega_4\}$. If one introduces
\begin{equation}\label{B.9}
\phi= \log(W/\tilde{W})\,,
\end{equation}
then $\phi$ is a collisional invariant in the sense that
\begin{equation}\label{B.10}
\phi_1+\phi_2=\phi_3+\phi_4\quad \textrm{on } \mathcal{D}_{\mathrm{c}}\,.
\end{equation}
Under some regularity on $\omega$ and for $\int_{\mathbb{T}^d} dk
|\phi(k)|<\infty$, it is proved in \cite{HS05} that the only
solutions to (\ref{B.10}) are
\begin{equation}\label{B.11}
\phi(k)=a+b\omega(k)\,.
\end{equation}

If $W$ is stationary and if $\int_{\mathbb{T}^d} dk | \log
\big(W/(1-W)\big)|<\infty$, then $W$ is necessarily of the form
\begin{equation}\label{B.11a}
W^-_{\beta,\mu}(k)=\big(\mathrm{e}^{\beta(\omega(k)-\mu)}+ 1\big)^{-1}
\end{equation}
for some $\beta,\mu\in\mathbb{R}$. In the limit $\beta\to\infty$,
one obtains
\begin{equation}\label{B.12}
W^-_{\infty,\mu}(k)=\mathbbm{1}(\omega(k)\leq \mu)
\end{equation}
and for $\beta\to-\infty$
\begin{equation}\label{B.13}
W^-_{-\infty,\mu}(k)=\mathbbm{1}(\omega(k)\geq \mu)\,.
\end{equation}
Although the corresponding collision invariant is not integrable,
one checks directly that $W^-_{\infty,\mu}$ and $W^-_{-\infty,\mu}$
are stationary solutions of (\ref{B.1}), (\ref{B.2}). Presumably our
list comprises all stationary solutions of the Boltzmann-Nordheim
equation.

Let us introduce the density, $\rho$, and energy, $\mathsf{e}$, of $W$ through
\begin{equation}\label{B.14}
\rho(W)=\int_{\mathbb{T}^d} dk W(k)\,,\quad \mathsf{e}(W)=
\int_{\mathbb{T}^d} dk \omega(k) W(k)\,.
\end{equation}
If one picks some $W$, then there is a unique pair
$\beta,\mu$ such that $\rho(W) = \rho(W_{\beta,\mu}^-)$, $\mathsf{e}(W) =
\mathsf{e}(W_{\beta,\mu}^-)$, provided one also admits the values
$\beta=\pm\infty$. Because of the conservation laws, this strongly
suggests that for the solution with this $W$ as initial datum it holds
\begin{equation}\label{B.17}
\lim_{t\to\infty} W(t)=W_{\beta,\mu}^-\,.
\end{equation}

On the mathematical side, besides the study of collision invariants,
the main focus so far is the existence and uniqueness of solutions.
We refer to the review \cite{EMV}. These authors consider the
momentum space $\mathbb{R}^3$ and dispersion
 $\omega(k)=k^2$, see Remark 5.3. To be
concise we quote one result from Dobeault \cite{Do}.\medskip\\
\textbf{Theorem B.1} \textit{Let
$d\geq 2$ and $\omega(k)=k^2$. Let $V$ be radial with
$\int_{\mathbb{R}^d} dk |k| |\widehat{V}(k)|^2<\infty$. For the
initial datum, $W$, we assume $W\in L^\infty(\mathbb{R}^d)$ and
$0\leq W\leq 1$. Then the integrated version of the
Boltzmann-Nordheim equation,
\begin{equation}\label{B.18}
W(t)=W+\int^t_0 ds \mathcal{C}(W(s))\,,
\end{equation}
has a unique solution in $C([0,\infty), L^\infty(\mathbb{R}^d))$,
which satisfies $0\leq W(t)\leq 1$ for all $t\geq 0$. If
$\rho(W)<\infty$, $\mathsf{e}(W)<\infty$, and $S(W)<\infty$, then it
holds, for all $t\geq 0$,}
\begin{eqnarray}\label{B.18a}
&&\hspace{-20pt} \int_{\mathbb{R}^d} dk W(k,t)=\rho(W)\,,\quad
\int_{\mathbb{R}^d} dk k W(k,t)=\int_{\mathbb{R}^d} dk  k
W(k)\,,\nonumber\\
&&\hspace{-20pt}\int_{\mathbb{R}^d} dk k^2
W(k,t)=\mathsf{e}(W)\,,\quad S(W(t))= S(W)+\int^t_0 ds
\sigma\big(W(s)\big)\,.
\end{eqnarray}

Apparently, the case of interest in our context has never been 
investigated. One difficulty results from the constraint due to
the energy $\delta$-function. On the other hand, since $\mathbb{T}^d$ is
compact, finite number, energy, and entropy holds automatically.
Under the Assumption A2 we expect Theorem B.1 still to be valid.

\subsection{Bosons}\label{B.2}

The Boltzmann-Nordheim equation reads, see (\ref{6a.4}),
(\ref{6a.5}),
\begin{equation}\label{B.19}
\frac{\partial}{\partial t} W(k,t)=\mathcal{C}(W(t))(k)\,,
\end{equation}
\begin{eqnarray}\label{B.19a}
&&\hspace{-53pt}\mathcal{C}(W)(k)= \pi \int_{(\mathbb{T}^d)^3} dk_2
dk_3 dk_4 \delta(k_1+k_2-k_3-k_4)\delta(\omega_1
+\omega_2-\omega_3-\omega_4)\nonumber\\
&&\hspace{15pt}|\widehat{V}(k_2-k_3)+ \widehat{V}(k_2-k_4)|^2
(\tilde{W}_1\tilde{W}_2 W_3 W_4-W_1 W_2\tilde{W}_3\tilde{W}_4)\,,
\end{eqnarray}
$\tilde{W}=1+W$. It has to be solved with initial data $W$ such that
$W\geq 0$, $\int dk W(k)<\infty$. Positivity is preserved in the
course of time and number and energy conservation holds as in
(\ref{B.3}).

According to the Bose statistics, the entropy per volume of a
quasifree state is given by
\begin{equation}\label{B.20}
S(W)=-\int_{\mathbb{T}^d} dk \big(W(k)\log W(k)-
\tilde{W}(k)\log\tilde{W}(k)\big)\,.
\end{equation}
Then (\ref{B.5}) to (\ref{B.11}) still hold, provided
$\tilde{W}=1+W$ and in (\ref{B.6}) $|\widehat{V}(k_2-k_3)-
\widehat{V}(k_2-k_4)|^2$ is substituted by $|\widehat{V}(k_2-k_3)+
\widehat{V}(k_2-k_4)|^2$.

If $W$ is stationary and $\int_{\mathbb{T}^d} dk |\log
W/(1+W)|<\infty$, then $W$ is necessarily of the form
\begin{equation}\label{B.21}
W^+_{\beta,\mu}(k)=\big(\mathrm{e}^{\beta(\omega(k)-\mu)}-1\big)^{-1}\,,
\end{equation}
where, in order to have $W^+_{\beta,\mu}\geq 0$, one is restricted
to
\begin{equation}\label{B.22}
\beta> 0\,,\quad \omega_\mathrm{min}\geq \mu \quad \mathrm{and}\quad
\beta< 0\,,\quad \mu\geq\omega_\mathrm{max}\,.
\end{equation}

The density and energy of a state $W$ is still given by
(\ref{B.14}). Clearly, the range of this map is
$D =\{(\rho,\mathsf{e})|0<\rho<\infty,\rho\omega_\mathrm{min}
<\mathsf{e}<\rho\omega_\mathrm{max},\}$.
However for $d \geq 3$ and under our assumptions on $\omega$, the map $(\beta,\mu)
\to (\rho(W^+_{\beta,\mu}), \mathsf{e}(W^+_{\beta,\mu}))$ has a
range $D_{\mathrm{nor}}$ which is a proper subset of $D$. Therefore,
for an initial $W$ with $(\rho(W),\mathsf{e}(W)) \in D \setminus 
D_{\mathrm{nor}}$ the solution of the Boltzmann-Nordheim 
equation has no limiting stationary state. Physically the 
excess mass condenses to a superfluid component, which is
reflected by a $\delta$-peak at $k=0$ in the Wigner function. 
Based on self-similar solutions it is argued that the 
condensate is nucleated at some finite time. 
We refer to the review  \cite{Pom}, see also \cite{ST97,Spo08}. 

On the mathematical side, the cubic nonlinearity of the
Boltzmann-Nordheim equation poses severe difficulties. Only the case of
momentum space $\mathbb{R}^3$, dispersion $\omega(k)=k^2/2$ has been
studied in detail and even then only for isotropic solutions which makes
$W(k,t)$ to depend on $|k|$ only. We use $\varepsilon$ as energy
variable, $\varepsilon=k^2/2$, and set $W(k,t)=f(\varepsilon,t)$.
In case $\widehat{V}(k) = \widehat{V}(0)=1/(2(2\pi)^3)$ the Boltzmann-Nordheim equation
for the energy distribution reads
\begin{equation}\label{B.25a}
\frac{\partial}{\partial t}
f(\varepsilon_1,t)=\mathcal{C}_\mathrm{r}\big(f(t)\big)(\varepsilon_1)
\end{equation}
with the collision operator
\begin{equation}\label{B.25b}
\mathcal{C}_\mathrm{r}
(f)(\varepsilon_1)=\int_{\{\varepsilon_3+\varepsilon_4\geq
\varepsilon_1\}} d\varepsilon_3 d\varepsilon_4
\frac{1}{\sqrt{\varepsilon_1}} \min
\big\{ \sqrt{\varepsilon_1},\sqrt{\varepsilon_2},\sqrt{\varepsilon_3},\sqrt{\varepsilon_4}\,
\big\}
(\tilde{f}_1\tilde{f}_2 f_3 f_4 -f_1 f_2 \tilde{f}_3 \tilde{f}_4)\,,
\end{equation}
see \cite{ST97}. Here
$\varepsilon_2=\varepsilon_3+\varepsilon_4-\varepsilon_1$,
$f_j=f(\varepsilon_j)$, $j=1,\ldots,4$, $\tilde{f}=1+f$. Lu
\cite{Lu04} considers Equation (\ref{B.25a}) in the weak form
\begin{equation}\label{B.25c}
\frac{d}{dt}\int^\infty_0 
\varphi(\varepsilon) f(\varepsilon,t)\sqrt{\varepsilon}d\varepsilon
= \int^\infty_0
\varphi(\varepsilon)\mathcal{C}_r\big(f(t)\big)(\varepsilon)
\sqrt{\varepsilon}d\varepsilon
\end{equation}
with test functions $\varphi\in C_{\mathrm{b}}^2(\mathbb{R}_+)$ and proves
that (\ref{B.25c}) remains meaningful even if $f(\varepsilon,t)\sqrt{\varepsilon}d\varepsilon$ is 
a positive
measure on $\mathbb{R}_+$. We denote such a measure by
$f(d\varepsilon,t)$. The possible roughness of $f(d\varepsilon,t)$ is 
balanced 
by rewriting the right hand side of (\ref{B.25c}) in such a way that the 
collisional difference $\varphi(\varepsilon_1) + \varphi(\varepsilon_2) - \varphi(\varepsilon_3)
-\varphi(\varepsilon_4)$ appears. For $t=0$ we assume finite mass and energy,
i.e., $\int^\infty_0f(d\varepsilon)<\infty$ and
$\int^\infty_0\varepsilon f(d\varepsilon)<\infty$.
Then (\ref{B.25c}) has a solution which conserves mass and energy.
The stationary measures of (\ref{B.25c}) are necessarily of the form
\begin{equation}\label{B.25d}
f_{\beta,\mu}(d\varepsilon)=(\mathrm{e}^{\beta(\varepsilon-\mu)}-1)^{-1}
\sqrt{\varepsilon} d\varepsilon+n_\mathrm{con}\delta(\varepsilon)d\varepsilon
\end{equation}
with either $\mu\leq 0$ and condensate density $n_\mathrm{con} = 0$ or $\mu=0$ and
$n_\mathrm{con} > 0$. The critical line dividing the normal fluid from the condensate
is given by $\mathsf{e}(\rho) = (\rho/\rho_0)^{-5/3}\mathsf{e}_0$
with $\rho_0 = \int_0^\infty d \varepsilon
\sqrt{\varepsilon}(e^{\beta\varepsilon} -1)^{-1}$, $\mathsf{e}_0 =\int_0^\infty d \varepsilon
\sqrt{\varepsilon}\varepsilon(e^{\beta\varepsilon} -1)^{-1}$.

For the particular case (\ref{B.25a}) one has some rigorous
information on the concentration. Let the initial data be given by a
density, i.e.,
$f(d\varepsilon)=f_{\mathrm{ac}}(\varepsilon)\sqrt{\varepsilon}
d\varepsilon$, such that
\begin{equation}\label{B.25e}
\rho=\rho(f)=\int^\infty_0
f_{\mathrm{ac}}(\varepsilon)\sqrt{\varepsilon}
d\varepsilon<\infty\,,\quad
\mathsf{e}=\mathsf{e}(f)=\int^\infty_0
\varepsilon f_{\mathrm{ac}}(\varepsilon)\sqrt{\varepsilon}
d\varepsilon<\infty\,.
\end{equation}
In general, the solution to (\ref{B.25c}) is then a measure,
$f(d\varepsilon,t)$, which can be uniquely decomposed into an
absolutely continuous and a singular part,
\begin{equation}\label{B.30}
f(d\varepsilon,t)=f_\mathrm{ac}(\varepsilon,t)\sqrt{\varepsilon}d\varepsilon
+ f_\mathrm{s}(d\varepsilon,t)\,.
\end{equation}
Now let $f_{\beta,\mu}$ be the equilibrium distribution
corresponding to $\rho(f)$, $\mathsf{e}(f)$, where $\mu=0$ in case
$n_{\mathrm{con}}  > 0$.  Lu \cite{Lu05} proves that, for
$\rho\leq\rho_\mathrm{c}$,
\begin{equation}\label{B.31}
\lim_{t\to\infty} \int^\infty_0
f_\mathrm{s}(d\varepsilon,t)=0\,,\quad \lim_{t\to\infty}
\int^\infty_0 
|f_\mathrm{ac}(\varepsilon,t)-f_{\beta,\mu}(\varepsilon)| 
\sqrt{\varepsilon}d\varepsilon=0\,.
\end{equation}
On the other hand for $\rho >\rho_\mathrm{c}$ one has
\begin{equation}\label{B.31a}
\lim_{t\to\infty} \int^\infty_0
|f_\mathrm{ac}(\varepsilon,t)-f_{\beta,0}(\varepsilon)|
\varepsilon\sqrt{\varepsilon}d\varepsilon =0
\end{equation}
and 
\begin{equation}\label{B.32}
\lim_{t\to\infty} \int_{\{\varepsilon|\,0 \leq \varepsilon^4\leq r(t)\}}
f(d\varepsilon,t)=\rho - \rho_\mathrm{c}\,,
\end{equation}
where $r(t)$ is the integral in (\ref{B.31a}). 
Thus for large times the solution develops a $\delta$-peak at
$\varepsilon=0$ with weight $n_{\mathrm{con}}$.

Whether such a concentration appears after some finite time or only
asymptotically, as $t\to\infty$, is left unanswered by the theorem.

\end{appendix}

\end{document}